\documentclass[aps,prd,twocolumn,a4paper,superscriptaddress,nofootinbib]{revtex4-1}
\usepackage{color}
\usepackage{graphicx}
\usepackage{hyperref}   
\usepackage{appendix}
\usepackage{epsfig}
\usepackage{epstopdf}
\usepackage{amsmath}
\usepackage{subfig}
\usepackage{comment}
\usepackage{bm}
\usepackage{ulem}
\usepackage{cancel}
\usepackage{slashed}
\usepackage[dvipsnames]{xcolor}
\usepackage{float}
\hypersetup{pdftex,colorlinks=true,linkcolor=blue,citecolor=magenta,menucolor=black,urlcolor=blue,filecolor=blue}

\begin{document}

\author{He-Xia Zhang}
\email{hxzhang@m.scnu.edu.cn}
\affiliation{Key Laboratory of Atomic and Subatomic Structure and Quantum Control (MOE), Guangdong Basic Research Center of Excellence for Structure and Fundamental Interactions of Matter, Institute of Quantum Matter, South China Normal University, Guangzhou 510006, China}
\affiliation{Guangdong-Hong Kong Joint Laboratory of Quantum Matter, Guangdong Provincial Key Laboratory of Nuclear Science, Southern Nuclear Science Computing Center, South China Normal University, Guangzhou 510006, China}

\author{Ke-Ming Shen}
\email{shen$\_$keming@ecut.edu.cn}
\affiliation{School of Science, East China University of Technology, Nanchang 330013, China}

\author{Yu-Xin Xiao}
\email{xiaoyx@mails.ccnu.edu.cn}
\affiliation{ Key Laboratory of Quark \& Lepton Physics (MOE) and Institute of 
		Particle Physics, Central China Normal University, Wuhan 430079, China}

\author{Ben-Wei Zhang}
  \email{bwzhang@mail.ccnu.edu.cn}
\affiliation{ Key Laboratory of Quark \& Lepton Physics (MOE) and Institute of 
		Particle Physics, Central China Normal University, Wuhan 430079, China}

\title{Impact of (magneto-)thermoelectric  effect on diffusion of conserved charges in hot and dense hadronic matter}

\begin{abstract}
We investigate the thermoelectric effect, which describes the generation of an electric field induced by temperature and conserved charge chemical potential gradients, in the hot and dense hadronic matter created in heavy-ion collisions.
Utilizing the Boltzmann kinetic theory within the repulsive mean-field hadron resonance gas model, we evaluate both the diffusion thermopower matrix and diffusion coefficient matrix for the baryon number ($B$), electric charge ($Q$), and strangeness ($S$).  
The Landau-Lifshitz choice for the rest frame of the fluid is enforced in the derivation.
We find that the thermoelectric effect hinders the diffusion processes of multiple conserved charges, particularly reducing the coupling between electric charge and baryon number (strangeness) in baryon (strangeness) diffusion. 
 Given that the repulsive mean-field interactions between hadrons have a significant effect on the diffusion thermopower matrix and diffusion coefficient matrix in the baryon-rich region, we extend the investigation to include the impact of magnetic fields, analyzing the magneto-thermoelectric effect on both the diffusion coefficient matrix and the Hall-like diffusion coefficient matrix. The sensitivities of the magnetic field-dependent diffusion thermopower matrix and magneto-thermoelectric modified diffusion coefficient matrix to the choices of various transverse conditions are also studied.
 
\end{abstract}

\maketitle

%-----------------------------------------------------------------------------------
\section{Introduction}

Relativistic heavy-ion collision experiments open up a unique portal for understanding the properties of strongly interacting matter under extreme temperatures.
A wealth of experimental data from the relativistic heavy-ion collider (RHIC)~\cite{BRAHMS:2004adc, PHOBOS:2004zne, PHENIX:2004vcz, STAR:2005gfr} at Brookhaven National Laboratory (BNL) and in the large hadron collider (LHC)~\cite{ALICE:2010khr, CMS:2011aqh, CMS:2011iwn, CMS:2012xss, ATLAS:2011ah, ALICE:2010yje} at the European Organization for Nuclear Research (CERN) have indicated that a new deconfined state of matter -- quark-gluon plasma (QGP) can be created. 
Meanwhile, quantum chromodynamics (QCD) serves as the fundamental theory of the strong interaction and the lattice QCD calculation has predicted a smooth crossover for QCD matter from a hadronic phase to a QGP phase can be realized as temperature ($T$) increases at the small or vanishing baryon chemical potential ($\mu_{B}$)~\cite{Aoki:2006we, Cheng:2006qk, Philipsen:2012nu}.
At large $\mu_{B}$,  calculations based on the low-energy QCD effective models, such as the (Polyakov-loop-) Nambu-Jona-Lasinio model~\cite{Nambu:1961tp, Hatsuda:1994pi, Fukushima:2008wg}, the (Polyakov-loop-) quark-meson model~\cite{Schaefer:2006ds, Schaefer:2007pw, Schaefer:2008hk, Schaefer:2011ex} have revealed that the QCD phase transition becomes first-order and terminates at a second-order critical endpoint (CEP), which has sparked long-sought debate without conclusive experimental evidence yet~\cite{Lacey:2014wqa}.  
Furthermore, the Beam Energy Scan (BES) program at RHIC~\cite{STAR:2010vob, Mohanty:2011nm} and ongoing experimental programs at the Facility for Antiproton and Ion Research (FAIR)~\cite{Tahir:2005zz, CBM:2016kpk} and the Nuclotron-based Ion Collider fAcility (NICA)~\cite{Friman, Kekelidze:2012zz}, are striving to unravel the properties of baryon-dense nuclear matter and search the potential signatures of the CEP in the QCD phase diagram.

In addition to the equilibrium QCD thermodynamical properties,  the medium's response to perturbations around equilibrium, which is encoded in transport coefficients, plays a crucial role in describing the evolution of bulk matter created in relativistic heavy-ion collisions.
The small shear viscosity to entropy density ratio ($\eta/s$) has been employed to successfully describe collective flow observables~\cite{Romatschke:2007mq, Kovtun:2004de, Heinz:2013th,Demir:2008tr}.
The bulk viscosity to entropy density ratio ($\zeta/s$) exhibits a novel behavior near critical temperature~\cite{Sasaki:2008fg, Karsch:2007jc,Meyer:2007dy}.
 Recently, electric conductivity has been utilized to extend the duration of the initial magnetic field generated in off-central heavy-ion collisions~\cite{Voronyuk:2011jd, Tuchin:2010gx, McLerran:2013hla}, and provide essential input for magnetohydrodynamic simulations~\cite{Roy:2015kma, Roy:2017yvg}.
On the other hand, the diffusion coefficient, which characterizes the medium's response to inhomogeneities in the number density and is largely overlooked at the top RHIC and LHC energies,  gains significance in the dynamical description of the evolution of low-energy heavy-ion collisions. Specifically, a
 diffusion coefficient matrix is specifically required to quantify the coupling among diffusion currents of various conserved charges.
This arises from the fact that  QCD matter constituents, such as hadrons and quarks, carry multiple quantum conserved numbers, including baryon number ($B$), electric charge ($Q$), and strangeness ($S$).  
The complete diffusion coefficient matrix in both the QGP and hadron gas has been calculated within the Boltzmann kinetic theory~\cite{Greif:2017byw, Fotakis:2019nbq, Fotakis:2021diq}. The associated results reveal that the off-diagonal matrix elements are comparable in magnitude to the diagonal elements, making them crucial in 
the hydrodynamic simulation of low-energy nuclear collisions. 
Recently, A. Das et. al. have explicitly imposed the Landau-Lifshitz frame into the previous derivations~\cite{Greif:2017byw, Fotakis:2019nbq}, and provided a unique expression in the hadron resonance gas model with and without excluded volume corrections~\cite{Das:2021bkz}. 
In this study, we explore how repulsive interactions, incorporated via a density-dependent mean-field potential in the repulsive mean-field hadron resonance gas (RMFHRG) model, impact the diffusion properties of hadronic matter.

The thermoelectric effect,  which facilitates the direct conversion of temperature differences into electric voltage and vice versa through a thermocouple, has been extensively explored across various disciplines, including material science, solid-state physics, and chemistry.
The most fundamental thermoelectric effect is the Seebeck effect, which describes the generation of an electric voltage caused by a temperature gradient within a conductive material. 
Intriguingly, heavy-ion collisions offer a unique platform to study the Seebeck effect in QCD matter. This is due to the notable temperature difference between the central region and peripheral region of the fireball. 
The Seebeck coefficient or thermopower, defined as the ratio of the induced electric field and the collinear temperature gradient ($S=\bm{E}/\bm{\nabla}T$) in the absence of electric current, has been estimated in both the partonic and hadronic phases of QCD matter~\cite{Zhang:2020efz, Das:2021qii, Zhang:2021xib, Kurian:2021zyb, Dey:2020sbm, Dey:2021crc, Khan:2022apd, Bhatt:2018ncr, Das:2020beh}. 
Unlike studies in condensed matter physics, the non-zero net conserved number is required to estimate the Seebeck coefficient in QCD matter. 
To our best knowledge, most estimations   have focused exclusively on scenarios where only net baryon density is nonzero,   neglecting the potential contributions of other conserved charges to the thermoelectric effect. However, it's important to note that the gradients of multiple conserved charge chemical thermal potentials ($\bm{\nabla} (\mu_q/T)$ with $q\in \{Q,B,S\}$) are also a source to generate an internal electric field, which is quantified by diffusion thermopower matrix, denoted as $M^{qQ}=\bm{E}/(T\bm{\nabla}(\mu_q/T))$, in the limit of zero electric current. 
Such thermoelectric effect involving multiple conserved charges has the potential to further influence the thermally spin Hall effect (TSHE) recently proposed in the heavy-ion collisions at BES energies~\cite{Liu:2020dxg, Fu:2022myl}. 
Given that the thermoelectric effect is highly related to the gradients of conserved charge chemical potentials and can theoretically impact the diffusion coefficient matrix, to the best of our knowledge, there have been no associated studies yet. This serves as the primary motivation for our research.

Considering that a partial magnetic field created in the off-central heavy-ion collisions can persist into the hadronic phase,
the motions of charged hadrons driven by  $\bm{\nabla}(T(\mu_q/T))$ 
 undergo deflection. This deflection results in a transverse or Hall-like electric field, viz, $\bm{E}\sim\bm{\nabla}(T(\mu_q/T))\times \bm{H}$ in the magnetic field. 
This phenomenon, known as the magneto-thermoelectric effect, is quantified by the Hall-like diffusion thermopower.
Similarly, additional Hall-type diffusion currents of conserved charges also arise in the magnetic field and are determined by the Hall-like diffusion coefficient matrix. 
Thus, the magneto-thermoelectric effect can affect both the magnetic field-dependent diffusion coefficient matrix and the Hall-like diffusion coefficient matrix.  
It is crucial to note that the formulation of the magneto-thermoelectric modified diffusion coefficient matrix depends on the choices of transverse conditions: 
1) all transverse gradients in net conserved charge densities vanish, 2) another transverse specific conserved charge diffusion current disappears apart from a zero transverse electric current.

This paper is organized as follows. 
In Sec.~\ref{sec:model}, we provide a brief overview of the ideal hadron resonance gas (IHRG) model and repulsive mean-field hadron resonance gas (RMFHRG) model. In Sec.~\ref{sec:coefficient}, we derive the 
general formulae for the diffusion thermopower and diffusion coefficient of conserved charges by solving the Boltzmann equation under relaxation time approximation in the framework of the RMFHRG model with and without magnetic field.  We present, for the first time, the expressions of the magneto-thermoelectric modified diffusion coefficient matrix 
under various transverse conditions.
Sec.~\ref{sec:results} delves into the impacts of RMF correction, baryon chemical potential, magnetic field, and (magneto-)thermoelectric effect on the (Hall-like-) diffusion coefficient matrix.  
We summarize our findings in Sec.~\ref{sec:summary}.

~
Throughout this paper, we adopt natural units with $c=\hbar=k_B=1$, and work in flat Minkowski space-time with metric tensor $g^{\mu\nu}=\mathrm{diag}(1,-1,-1,-1)$, thus the fluid velocity satisfies $u^\mu u_{\mu}=1$. The tensor $\Delta^{\mu\nu}=g^{\mu\nu}-u^\mu u^{\nu}$ is the projection operator onto the three-dimensional subspace orthogonal to $u^{\mu}$. In the local rest frame, $u^\mu=(1,\bm{0})$, the projector $\Delta^{\mu\nu}$ has the form: 
$ \Delta_{ \mu\nu}=\Delta^{\mu\nu}_{}=\mathrm{diag}(0,-1,-1,-1),~ \Delta^\mu_{\nu}=\mathrm{diag}(0,1,1,1)$. The projection of any four-vector $A^\mu=(A^0,\bm{A})$ onto the three-dimensional subspace orthogonal to $u^\mu$ is defined as $A^{\langle\mu\rangle}\equiv \Delta^\mu_{\nu}A^\nu$.
The four-derivative is decomposed as $\partial^\mu\equiv \nabla^\mu+u^{\mu}D $, where $D=u^\mu \partial_{\mu}$ and  $\nabla_\nu=\Delta^{\alpha}_{\nu}\partial_{\alpha}$ denote the time derivative and spatial gradient operator in the local rest frame, respectively. In the local rest frame, we have $\nabla^\mu\equiv (0,-\bm{ \nabla})$.

 \section{Model description}\label{sec:model}
The hadron resonance gas (HRG) model~\cite{Cleymans:1999st, Becattini:2000jw} is a simplistic thermal statistical model that successfully describes the low-temperature hadronic phase of QCD at chemical freeze-out. In the IHRG model, the attractive interactions between hadrons are implicitly accounted for by including all the resonances with zero width, while the repulsive interactions among hadrons, which are already known from nucleon-nucleon scattering experiments, are missed.
 Consequently, several extensions of the IHRG model have emerged, such as excluded volume HRG model~\cite{Rischke:1991ke, Andronic:2012ut, Kadam:2015xsa}, van der Waals HRG model~\cite{Vovchenko:2016rkn, Vovchenko:2017zpj} and repulsive mean-field HRG model~\cite{Pal:2023zzp, Huovinen:2017ogf, Kadam:2019peo, Pal:2020ucy}. These extensions aim to provide a more precise fit to various thermodynamic observables derived from lattice QCD simulations.
In this work, we utilize the repulsive mean-field hadron resonance gas (RMFHRG) model to describe the repulsive interactions between the hadrons.  %the repulsive interactions between the hadrons are incorporated via a mean-field approach, and this extended version of the IHRG model is the so-called repulsive mean-field hadron resonance gas (RMFHRG) model. 

%-----------------------------------------------------------------------------------
\subsection{Ideal hadron resonance gas (IHRG) model}
In the IHRG model, the partition function, containing all relevant degrees of freedom of the confined QCD phase, is the starting point for deriving thermodynamic observables. The logarithm of the total partition function in the grand canonical ensemble is given as 
\begin{eqnarray}
\ln Z^{id}=\sum_{a}\ln Z_a^{id} (T,\mu_a,m_a),
\end{eqnarray}
where logarithm of the partition function for hadron species $a$ is given by
\begin{eqnarray}
\ln Z_a^{id}=\pm V\sum_{a} \int d\Gamma_{a}\ln [1\pm e^{-\beta(\epsilon^0_a-\mu_a)}].
\end{eqnarray}
Here, the superscript ``$id$" represents the ideal  gas and $V$ is the system volume, we use the notation $d\Gamma_{a}=d_{a}d^{3}p_{a}/(2\pi)^{3}$, where $d_a$ is the spin degeneracy of hadron species $a$.  $\beta=1/T$ is the inverse temperature of the system. The front upper and lower signs correspond to (anti-)fermions and bosons, respectively.  $\epsilon^0_a=\sqrt{\bm{p}_a^2+m_a^2}$ is the energy of hadron species $a$ with mass $m_a$. The chemical potential of hadron species $a$ is defined as $\mu_a(\{\mu_q\})%=\sum_{q\in \{Q,B,S\}} q_a\mu_q
=B_a\mu_B+Q_a\mu_Q+S_a\mu_S$, where $\{\mu_q\}\equiv\{\mu_{B},\mu_{Q},\mu_S\}$ are the baryon, electric and strangeness chemical potentials, respectively, and  $B_a$, $Q_a$ and $S_a$ 
are the corresponding quantum numbers for particle species $a$. 
Therefore,  the ideal thermodynamics including total pressure $(P^{id})$, total energy density $(\mathcal{E}^{id})$, and total number density $(\rho^{id})$ in the IHRG model can be obtained as follows.
\begin{eqnarray}
P^{id}&=&\sum_{a}P_a^{id}=\frac{\partial \ln Z^{id} }{\beta \partial V}=\sum_{a}\int d\Gamma_{a}\frac{ \bm{p}_a^2}{3\epsilon^0_a}f_a^{id}\label{eq:TMeq1},\\
\mathcal{E}^{id}&=&\sum_{a}\mathcal{E}_a^{id}=-\frac{1}{V}\frac{\partial \ln Z^{id} }{\partial\beta }=\sum_{a}\int d\Gamma_{a} \epsilon^0_af_a^{id},\label{eq:TMeq2}\\
\rho^{id}&=&\sum_{a}\rho_a^{id}=\frac{T}{V}\frac{\partial \ln Z^{id} }{\partial \mu_a}=\sum_{a}\int d\Gamma_{a}f_a^{id}.\label{eq:TMeq3}
\end{eqnarray} 
Here, $f_a^{id}$ is the thermal equilibrium distribution function of particle species $a$ in the IHRG model. It is expressed as 
\begin{eqnarray}
f_a^{id}(T,\mu_a)=[\exp[\left(\epsilon_a^0-\mu_a\right)\beta]\pm 1]^{-1},
\end{eqnarray}
where the signs $\pm$ correspond to the Fermi-Dirac and Bose-Einstein statistics, respectively.

%-----------------------------------------------------------------------------------
\subsection{Repulsive mean-field hadron resonance gas (RMFHRG) model}
The RMFHRG model is a 
extension of the IHRG model that includes short-range repulsive interactions between hadrons via a mean-field approach. In this model, the single-particle energy is modified as~\cite{Olive:1980dy}
\begin{eqnarray}
\widetilde{\epsilon}_a=\epsilon^0_a+U_a,
\end{eqnarray}
where $U_a$ is the potential describing the repulsive interactions between hadrons, and acts as an additional chemical potential.  In the RMFHRG model, the repulsive interactions only among meson-meson pairs, baryon-baryon pairs, and antibaryon-antibaryon pairs are considered~\cite{Kadam:2019peo}. The mean-field potentials for (anti-)baryons and mesons are defined as \cite{Olive:1980dy,Kadam:2019peo}
\begin{equation}
U_{a\in\{B,\bar{ B}\}}(\rho_{B,\bar{B}})=K_B\rho_{B,\bar{B}},~ U_{a\in\{M\}}(\rho_{M})=K_M\rho_{M},
\end{equation}
with subscripts $B,~\bar{B}$, $M$ denoting baryons, antibaryons, and mesons, respectively, $\rho_{B,\bar{ B},M}$ are the respective total number densities. Two phenomenological parameters $K_M$ and $K_B$ are introduced to scale the repulsive interaction strength among the mesons and (anti-)baryons, respectively. Accordingly, the logarithm of the total partition function in the RMFHRG model can be expressed as 
\begin{eqnarray}\label{eq:ZRMF}
\ln Z^{\mathrm{RMF}}&=&\pm V
\sum_a \int d\Gamma_{a}\ln[1\pm e^{-\beta(\epsilon^0_a-\mu_a^*)}]\nonumber\\ &&-V\beta\chi_{a\in\{M,B,\bar{B}\}}(\rho_{M,B,\bar{B}}) 
\end{eqnarray}
where the effective chemical potential of hadron species $a$ is defined as $\mu_a^*=\sum_qq_a\mu_q-K_{B,\bar{B}}\rho_{B,\bar{B}}$ for (anti-)baryons, and  $\mu_a^*=\sum_qq_a\mu_q-K_{M}\rho_{M}$ for mesons. In the RMFHRG model,
$\rho_{B,\bar{B},M}$ are calculated as follows:
\begin{equation}\label{eq:rho}
\rho_{B,\bar{B},M}=\sum\limits_{a\in\{B,\bar{B},M\}}\int d\Gamma_{a}\bar{f}_a^0.
\end{equation}
Here, $\bar{f}_a^0$ represents the thermal equilibrium distribution function of particle species $a$ in the RMFHRG model. It is given by:
\begin{equation}
\bar{ f}_a^0(T,\mu_a)= f_a^{id}(T,\mu_a^*)=[\exp([\epsilon^0_a- \mu_a^*]\beta)\pm 1]^{-1}.
\end{equation}
In Eq.~(\ref{eq:ZRMF}), $\chi_a$ is an additional correction factor to avoid the double counting of the mean-field potential and to ensure the correct number density per particle $\rho_a=\frac{T}{V}\frac{\partial \ln Z^{\mathrm{RMF}}}{\partial \mu_a}$ (or the correct energy density per particle $\mathcal{E}_a=\partial \mathcal{E}/\partial \rho_a$).  Assuming that hadron species $a$ is a baryon or antibaryon, its number density takes the following form:
\begin{eqnarray}
\rho_a=\rho_a-\rho_a\left(\frac{\partial U_a}{\partial \rho_{B,\bar{ B}}}\frac{\partial \rho_{B,\bar{ B}}}{\partial \mu_a}\right)-\frac{\partial \chi_a}{\partial \rho_{B,\bar{B}}}\frac{\partial \rho_{B,\bar{ B}}}{\partial \mu_a}.
\end{eqnarray}
Accordingly, the expression of $\chi_a$ can be obtained as 
\begin{eqnarray}
\chi_{a\in\{ B,\bar{B}\}}(\rho_{B,\bar{B}})=-\frac{1}{2}K_{B}\rho_{B,\bar{B}}^2.
\end{eqnarray}
Similarly, if hadron species $a$ is a meson, $\chi_{a\in\{ M\}}(\rho_{M})=-\frac{1}{2}K_{M}\rho_{M}^2$.
%By utilizing Eqs.~(\ref{eq:TMeq1}-\ref{eq:TMeq2}) along with Eq.~(\ref{eq:ZRMF}), we can derive the expressions for pressure and energy density in the RMFHRG model for baryons, antibaryons, and mesons.
By utilizing Eqs.~(\ref{eq:TMeq1}-\ref{eq:TMeq2}) along with Eq.~(\ref{eq:ZRMF}), we can derive the pressure and energy density for baryons, antibaryons, and mesons in the RMFHRG model, which are expressed respectively as 
\begin{align}
{P}_{B,\bar{B},M}&=\sum_{a\in\{B,\bar{ B},M\}}P_a^{id}(T,\mu_a^*)-\chi_{B,\bar{B},M}(\rho_{B,\bar{B},M}),\\
{\mathcal{E}}_{B,\bar{ B},M}&=\sum_{a\in\{B,\bar{ B},M\}}\mathcal{E}_a^{id}(T,\mu_a^*)+\chi_{B,\bar{B},M}(\rho_{B,\bar{B},M}).
\end{align}
The total pressure and total energy density in the RMFHRG model are given by ${ P}={ P}_{B}+{ P}_{\bar{ B}}+{ P}_{M}$ and ${\mathcal{E}}={\mathcal{E}}_{B}+{\mathcal{E}}_{\bar{ B}}+{\mathcal{E}}_{M}$, respectively. Compared to the IHRG model, the RMFHRG model incorporates an additional term in both pressure and energy density,   ensuring the thermodynamic consistency. In the present RMFHRG model, all distinct (anti-)baryons or mesons are assigned a uniform repulsive interaction strength. Specifically,  $K_B=0.45$~$\mathrm{GeV\cdot fm^3}$ and $K_M=0.05$~$\mathrm{GeV\cdot fm^3}$ are adopted to improve the agreement with the thermodynamic quantities obtained from the lattice QCD simulations at zero and finite baryon densities~\cite{Kadam:2019peo, Pal:2020ucy}.

%-----------------------------------------------------------------------------------
\section{Thermoelectric coefficients and diffusion coefficients of conserved charges in Boltzmann kinetic theory}\label{sec:coefficient}
\subsection{Formalism}
It is effective to calculate the transport coefficients of hadronic matter within the kinetic theory framework. The evolution of the single-particle phase-space distribution function $f_a(x,p_a)$ can be described by the Boltzmann equation within the covariant formalism~\cite{SMASH:2016zqf},
\begin{equation}\label{eq:Boltzmann_eq}
p_a^{\mu}\partial_{\mu}f_a(x,p_a)  + m_aK^{\mu}\partial_\mu^{(p)}f_a(x,p_a)= \mathcal{C}_a [f_a],
\end{equation}
where $p_a^{\mu}=(\epsilon^0_a, \bm{p}_a)$ represents the four-momentum of the particle species $a$, $K^\mu$ is the four-force experienced by individual particle, and $\mathcal{C}_a[f_a]$ denotes the collision term. When a particle is subjected to an electromagnetic field force, then $K_a^\mu=-\epsilon^0_a\partial^\mu U_a(x)/m_a +\frac{Q_a}{m_a}p_{a\nu}F^{\mu\nu}$. Here, $U_a$ refers to the repulsive mean-field potential among hadrons, and  
$F^{\mu\nu}$ is the electromagnetic field-strength tensor. 
This tensor is defined as: $F^{\mu\nu}\equiv E^\mu u^\nu-E^\nu u^\mu+\frac{1}{2}\epsilon^{\mu\nu\alpha\beta}(u_{\alpha}H_{\beta}-u_\beta H_\alpha)$, where $\epsilon^{\mu\nu\alpha\beta}$ is the totally anti-symmetric Levi-Civita tensor. The four-vectors $E^\mu\equiv F^{\mu\nu}u_{\nu}$ and $H^\mu=\epsilon^{\mu\nu\alpha\beta}F_{\nu\alpha}u_{\beta}/2$ are nothing but the electric and magnetic fields measured in the frame where the fluid moves with a velocity $u^\mu$. Both $E^\mu=(0,\bm{E})$ and $H^\mu==(0,\bm{H})$ are  space like, satisfying $E^\mu u_\mu=0, H^\mu u_{\mu}=0$.  They can be normalized as $E^\mu E_{\mu}=-E^2,~H^\mu H_{\mu}=-H^2$, where $E\equiv|\bm{E}|$ and $H\equiv|\bm{H}|$. Note that in this study, the electric field is induced by gradients of conserved charge densities rather than a decaying external magnetic field.
	
Considering that the system is slightly deviated from the local equilibrium, the phase space  distribution function for hadron species $a$ can be formulated  as:
\begin{equation}\label{f_phi}
f_a=\bar{f}_a^0(1+\phi_a),
\end{equation}
where the deviation function $|\phi_a|\ll1$. Here, $\bar{f}_a^0$ represents the local equilibrium  distribution function within the RMFHRG model, given by
\begin{equation}\label{eq:fbar}
\bar{f}_a^0=[\exp((p_a^\mu u_\mu+U_a)\beta-\sum_{q}q_a\alpha_q)+1]^{-1},
\end{equation}
with $\alpha_{q}\equiv\mu_q\beta$ denoting the chemical thermal potential of conserved charge $q$.

To solve Eq.~(\ref{eq:Boltzmann_eq}), the  deviation function $\phi_a$ is assumed have the following linear combination form:
\begin{equation}\label{eq:phi_1}
\phi_a\approx -\sum_q\mathcal{B}_a^qp_a^\mu\nabla_\mu \alpha_q-
\mathcal{G}_ap_a^\mu\beta E_\mu+\dots,
\end{equation}
where $\mathcal{B}_a^q$ and $\mathcal{G}_a$ are unknown functions with respect to momentum $p_a$.  For the binary inelastic or reactive collisions $a(p_a)+b(p_b')\to c(p_c'')+d(p_d''')$, the initial particle species ($a,b$) are allowed to be different from final particles species ($c,d$),  such  that 
the collision term in Eq.~(\ref{eq:Boltzmann_eq}) reads as~\cite{Groot1980,Chakraborty:2010fr, Albright:2015fpa} 
\begin{align}\label{eq:Ca}
\mathcal{C}_a=&\frac{1}{2}\sum_{b,c,d}\int d\Gamma_{b}'d\Gamma_{c}''d\Gamma_{d}''' \big[W_{ab|cd}(p_a,p_b'|p_c'',p_d''')\nonumber\\
&\times f_c'' f_d'''
(1+ t_af_a)
(1+t_b f_b')-W_{cd|ab}(p_c'',p_d'''|p_a,p_b')\nonumber\\
&\times f_a f_b'(1+ t_c f_c'')(1 +t_d f_d''')\big],
\end{align}
where $(1+t_a f_a)$ with $t_a=\pm $ correspond to  Bose enhancement factor and Pauli blocking factor, respectively. Here, $W_{ab|cd}(p_a,p_b'|p_c'',p_d''')$ is collisional transition rate, in the absence of a reaction threshold,  it satisfies the detailed balance property~\cite{Groot1980} 
 \begin{equation}
    W_{ab|cd}(p_a,p_b'|p_c'',p_d''')=
    W_{cd| ab}(p_c'',p_d'''|p_a,p_b').
\end{equation} 
Futhermore, the transition rate  remains  invariant under the interchange of momenta of incoming or outgoing particles:  $W_{ab|cd}(p_a,p_b'|p_c'',p_d''')=W_{ba|cd}(p_b',p_a|p_c'',p_d''')=W_{ab|dc}(p_a,p_b'|p_d''',p_c'')=W_{ba|dc}(p_b',p_a|p_c'',p_d''')$. 
 The prefactor $1/2$ in Eq.~(\ref{eq:Ca}) is added  to correct the double counting from the symmetry under the exchange of the momenta of the final state $p_c''$ and $p_d''$.  By utilizing the detailed balance condition, for example, $a+b\leftrightarrow c+d$ gives
$\bar{f}_c^{0''} \bar{f}_d^{0'''}(1+t_a\bar{f}^0_a)(1+t_b\bar{f}_b^{0'})=\bar{f}^0_a \bar{f}_b^{0'}(1+t_c\bar{f}_c^{0''})(1+t_d \bar{f}_d^{0'''})$, then the collision term is computed as
\begin{align}
    \mathcal{C}_a
=&
\frac{1}{2}\int d\Gamma_b'd\Gamma_c''d\Gamma_d'''
W_{ab|cd}(p_a,p_b'|p_c'',p_d''')\nonumber\\
&\times\bigl\{\bar{f}_a^0\bar{f}_b^{0'}[(1+t_d\bar{f}_d^{0'''})\phi_c''+(1+t_c\bar{f}_c^{0''})\phi_d''']\nonumber\\
&-\bar{f}_c^{0''}\bar{f}_d^{0'''}[(1+t_b\bar{f}_b^{0'})\phi_a+(1+t_a \bar{f}_a^0)\phi_b']\bigr\}\\
=&-\frac{1}{2} \int d\Gamma_b'd\Gamma_c''d\Gamma_d'''
W_{ab|cd}(p_a, p_b'|p_c'',p_d''')\nonumber\\
&\times\frac{\bar{f}_a^0 \bar{f}_b^{0'}(1 +t_c \bar{f}_c^{0''})}{(1+t_a\bar{f}^0_a)} (1+t_d \bar{f}_d^{0'''})\phi_a.
\end{align}
Here, we have assumed that particle species $a$ is slightly out of equilibrium ($\phi_a\neq0$), while all other particles are in equilibrium ($\phi_b'=\phi_c''=\phi_d'''=0$). 
In this study, our focus is solely on elastic binary collisions.  The transition rate in the elastic limit is defined as~\cite{Groot1980}: 
\begin{align}
 W_{ab|cd}=\gamma_{ab}(\delta_{ac}\delta_{bd}+\delta_{ad}\delta_{bc})W_{ab},
\end{align}
where $\gamma_{ab}$ has been inserted to guarantee that $W_{ab}$ represents the transition rate both for the case of identical particle species ($a=b$) and particles of different species ($a\neq b$). Accordingly, 
the collision term for the elastic binary process $a(p_a)+b(p_b')\to a(p_a'')+b(p_b''')$ is formulated as 
\begin{align}\label{eq:solution}
\mathcal{C}_a=&
- \sum_b\gamma_{ab}\int d\Gamma_b'd\Gamma_a''d\Gamma_b'''
W_{ab}(p_a,p_b'|p_a'',p_b''')\nonumber\\
&\times\frac{\bar{f}_a^0 \bar{f}_b^{0'}(1+t_a \bar{f}_a^{0''})}{(1+t_a\bar{f}^0_a)} (1+t_b \bar{f}_b^{0'''})\phi_a.
\end{align}

 We shall now consider the collision term in Eq.~(\ref{eq:Boltzmann_eq}) using a simple and popular approximation, known as the relaxation time approximation (RTA)~\cite{Anderson}. Under the RTA, the collision term takes the form:
\begin{equation}\label{eq:RTA}
-\frac{\epsilon^0_a \bar{f}_a^0\phi_a}{\tau_a}=\mathcal{C}_a=-\frac{\epsilon^0_a\delta f_a}{\tau_a}.
\end{equation}
Here, $\tau_a$ denotes the relaxation time of hadron species $a$, describing how fast the system reaches the equilibrium again. The term $\delta f_a=f_a- \bar{f}^0_a$ is a perturbation term. Then, the energy-momentum tensor $T^{\mu\nu}$ and the net conserved charge four-current $N_q^\mu$ can be expressed in terms of the phase space distribution function as follows:
\begin{eqnarray}
T^{\mu\nu}&=&\sum_a\int d\Gamma_{a}\frac{p_a^{*\mu} p_a^\nu}{\epsilon^0_a}f_a+g^{\mu\nu}\chi_a,\\
 N_q^\mu&=&\sum_a q_a\int d\Gamma_{a}\frac{p_a^\mu}{\epsilon_a^0} f_a,
\end{eqnarray}
where $p_a^{*\mu}=p_a^\mu +U_a^\mu$. In a non-equilibrium system, the energy-momentum diffusion four-current and the conserved charge diffusion four-current are defined as 
 $W^\mu\equiv\Delta^{\mu}_{\nu}T^{\alpha\nu}u_{\alpha}$ and $V^\mu_{q}\equiv\Delta^\mu_{\nu}N_q^\nu$, respectively \cite{Fotakis:2022usk,Molnar:2016vvu}.  They can also be expressed as: 
\begin{align} 
W^\mu&
=\sum_a\int d\Gamma_{a}\frac{\Delta^{\mu}_{\nu}p_a^\nu (\epsilon_a^0+U_a) }{\epsilon_a^0} \bar{f}_a^0 \phi_a,\label{eq:deltaT0i}\\
V^\mu_{q}&
=\sum_{a}q_a\int d\Gamma_{a}\frac{\Delta^\mu_{\nu} p_a^\nu}{\epsilon^0_a}\bar{f}_a^0\phi_a.\label{eq:deltaV}
\end{align}
In ideal hydrodynamics, the fluid four-velocity is determined because the energy and conserved charge number currents are parallel to each other. The local rest frame of the fluid is then defined by the requirement that these currents vanish identically.  However, in dissipative hydrodynamics, the energy flow and charge number flow are separate, leading to a non-unique definition of the fluid four-velocity~\cite{DerradideSouza:2015kpt}.
There are two natural choices for fixing the local rest frame of fluid: Eckart frame  (or conserved charge frame) and Landau-Lifshitz frame (or energy frame). In the Eckart frame, the fluid velocity is parallel to one of the conserved charge currents, demanding the overall diffusion current of that conserved charge to be zero. However, in low-energy heavy-ion collisions, there are multiple conserved charges and are not necessarily non-vanishing in all regions of space-time, therefore, the definition of Eckart frame may not be suitable~\cite{Fotakis:2022usk}.  On the other hand, in the Landau-Lifshitz frame, the fluid velocity is parallel to the energy flow, requiring the total energy-momentum diffusion current to vanish in the local rest frame. 
The Landau-Lifshitz frame is our choice for the local rest frame of fluid. 

Upon substitution of Eq.~(\ref{eq:RTA}) into the right-hand side of Eq.~(\ref{eq:Boltzmann_eq}), we can compute the perturbation term $\delta f_a$ for the first-order gradient expansion as follows:
\begin{align}\label{eq:deltaf0}
\delta f_a=&
\frac{\tau_a}{\epsilon_a^0}
\bar{f}_a^0(1-\bar{f}_a^0)\bigg[p_a^\mu p_a^\nu\beta(u_{\mu }Du_{\nu}+\nabla_\mu u_{\nu})\nonumber\\
&+p_a^\mu (p_a\cdot u+U_a)(u_{\mu}D\beta +\nabla_\mu \beta)\nonumber\\
&-p_a^\mu \sum_q q_a(u_{\mu}D\alpha_{q} +\nabla_\mu \alpha_q)\nonumber\\
&+p_a^\mu\beta (u_{\mu}DU_a +\nabla_\mu U_a)
-p_a^\mu\beta (u_{\mu}DU_a +\nabla_\mu U_a)
 \nonumber\\
 &+Q_a\beta p_{a\nu} F^{\mu\nu} u_{\mu}\bigg].
\end{align}
In ideal hydrodynamics, taking the projection of $\partial_{\mu }T^{\mu\nu}=0$ along the direction orthogonal to  $u_\nu$, one gets $u_\nu \partial_\mu T^{\mu\nu}=u_{\nu}[D {\mathcal{E} u^{\nu}}+({\mathcal{E}}+{ P})\theta u^{\nu}+(\mathcal{E}+P)Du^{\nu}-\nabla^\nu P]=0$, where $\theta$ is the expansion rate.
Due to $u_{\nu}u^{\nu}=1$, we have $u_{\nu}\partial_\mu u^{\nu}=0$. Then we can derive $Du_\mu=\frac{1}{\omega}\nabla_\mu{ P}$  with $\omega$ being enthalpy density. Recalling the Gibbs-Duhem relation, $dP=sdT+\sum_q n_q d\mu_q$, one arrives at: $\nabla_\mu P=-\beta^{-1}\omega \nabla_{\mu}\beta+\beta^{-1}\sum_qn_q\nabla_{\mu}\alpha_q$, where $n_q$ is conserved net charge density. By invoking  momentum conservation $\nabla_\mu P=0$, we ultimately obtain: $\nabla_{\mu}\beta=\sum_q\frac{n_q}{\omega}\nabla_\mu\alpha_q$.

\subsection{For vanishing  magnetic field}
In Eq.~(\ref{eq:deltaf0}), we first consider only  spatial-dependent gradient terms and neglect the magnetic field effect,  $\delta f_a$ can be simplified to: 
\begin{align}
\delta f_a=&\frac{\tau_a}{\epsilon_a^0} \bar{f}_a^0(1-\bar{f}_a^0)\bigg[p_a^\mu p_a^\nu\beta\nabla_\mu u_{\nu}-Q_a\beta p_{a\nu}E^\nu\nonumber\\
&+p_a^\mu\sum_q\left((p_a\cdot u+U_a)\frac{n_q}{\omega}-q_a\right)\nabla_\mu \alpha_q\bigg].
\end{align}
We only retain the terms related to the conserved charge diffusion current, the above equation can be further reduced as
\begin{eqnarray}\label{eq:deltaf-zeroB}
\delta f_a\simeq
-\tau_{a}\frac{p_{a}^\mu}{\epsilon_a^0} F_a Q_aE_\mu+\sum_q\tau_{a}\frac{p_a^\mu}{\epsilon_a^0} H_a^q\nabla_\mu\alpha_{q}.
\end{eqnarray}
The functions of  $F_a$ and $H_a^q$ in Eq.~(\ref{eq:deltaf-zeroB}) are defined as follows:
\begin{align}
F_a=&\beta \bar{f}_a^0(1\pm\bar{f}_a^{0}),\\
H^q_a=&\left(\frac{n_q}{\omega}(\epsilon^0_a+U_a)-q_a\right)\bar{f}_a^{0}(1\pm\bar{f}_a^{0}).
\end{align}

In the Landau-Lifshitz frame
condition, the  the total energy-momentum diffusion current is required to vanish, i.e., $W^{\mu}=0$. By inserting  Eq.~(\ref{eq:phi_1}) into Eq.~(\ref{eq:deltaT0i}), one gets
\begin{equation}\label{eq:LLcondition}
\sum_{a}\int d\Gamma_{a} \widetilde{\epsilon}_a\frac{p_a^{\langle\mu\rangle}p_a^{\langle\nu\rangle}}{\epsilon_a^0}\left[\sum_{q}\mathcal{B}_a^q\nabla_\nu\alpha_q+\mathcal{G}_a\beta E_\nu\right] \bar{f}_a^0
=0.
\end{equation}
Given  that we have the particular solutions  $\mathcal{B}_{a,\mathrm{part}}^q$, $\mathcal{G}_{a,\mathrm{part}}$,
other solutions can be expressed as   $\mathcal{B}_a^q=\mathcal{B}_{a,\mathrm{part}}^q-b^q$ and  $\mathcal{G}_a=\mathcal{G}_{a,\mathrm{part}}-g$, respectively, where $b^q$ and $g$ are the constants independent of particle species $a$. Inserting these expressions into Eq.~(\ref{eq:LLcondition}) to determine the uniqueness of solutions, we arrive at:
\begin{align}
&\sum\limits_{a}\int d\Gamma_{a}\widetilde{\epsilon}_a\frac{p_a^{\langle\mu\rangle}p_a^{\langle\nu\rangle}}{\epsilon_a^0}\left[\sum_{q}\mathcal{B}_{a,\mathrm{part}}^q\nabla_\nu\alpha_q+\mathcal{G}_{a,\mathrm{part}}\beta {E}_\nu\right] \bar{f}_a^0\nonumber\\
&=\sum\limits_{a}\int d\Gamma_{a}\widetilde{\epsilon}_a\frac{p_a^{\langle\mu\rangle}p_a^{\langle\nu\rangle}}{\epsilon_a^0}\left[\sum_{q}b^q\nabla_\nu\alpha_q+g\beta E_\nu\right] \bar{f}_a^0.
\end{align}
Employing the identity $3T\omega=\sum\limits_a\int d\Gamma_{a}(\epsilon_a^0+U_a)\frac{\bm{p}_a^2}{\epsilon_a^0}\bar{f}_a^0$,
and comparing the coefficients of $\nabla_\nu\alpha_q$ and $\beta E_\nu$, we get
\begin{eqnarray}
b^q&=&\frac{1}{3T\omega}\sum_{a}\int d\Gamma_{a}(\epsilon_a^0+U_a)\frac{\bm{p}_a^2}{\epsilon_a^0}\mathcal{B}_{a,\mathrm{part}}^q \bar{f}_a^0,\label{eq:b^q}\\
g&=&\frac{1}{3T\omega}\sum_a\int d\Gamma_{a} (\epsilon_a^0+U_a)\frac{\bm{p}_a^2}{\epsilon_a^0}\mathcal{G}_{a,\mathrm{part}} \bar{f}_a^0.\label{eq:g}
\end{eqnarray}
Inserting Eq.~(\ref{eq:phi_1}) into Eq.~(\ref{eq:deltaV}),
and utilizing Eqs.~(\ref{eq:b^q}-\ref{eq:g}), the diffusion current of conserved charge $q'$ can be written as 
\begin{align}
    V^\mu_{q'}=&\sum_{a}q'_a\int d\Gamma_{a}\frac{ p_a^{\langle\mu\rangle} p_a^{\langle\nu\rangle}}{\epsilon^0_a} \bigg[-\sum_q(\mathcal{B}_{a,\mathrm{part}}^q-b^q)\nabla_\nu\alpha_q\nonumber\\
    &-(\mathcal{G}_{a,\mathrm{part}}-g)\beta  E_\nu\bigg] \bar{f}_a^0\\
    =&-\sum_{a}q'_a\int d\Gamma_{a}\frac{ p_a^{\langle\mu\rangle} p_a^{\langle\nu\rangle}}{\epsilon^0_a} \bigg[\sum_q\mathcal{B}_{a,\mathrm{part}}^q\nabla_\nu\alpha_q\nonumber\\
    +&\mathcal{G}_{a,\mathrm{part}}\beta  E_\nu\bigg] \bar{f}_a^0 -n_q'T\sum_b b^q\nabla^\mu \alpha_q-g n_{q}'E^\mu,
    \label{eq:Vq'_part2}
\end{align}
where the orthogonality relation of $p_a^{\langle\mu\rangle} $: $p_a^{\langle\mu\rangle} p_a^{\langle\nu\rangle}=\frac{\Delta ^{\mu\nu}}{3}p_a^{\langle\gamma\rangle}p_{a\langle\gamma\rangle}=-\frac{\bm{p}_a^2}{3}\Delta^{\mu\nu}$ and the identity  $\sum\limits_{a}\int q'_a d\Gamma_{a}\frac{\bm{p}_a^2}{3\epsilon^0_a}\bar{f}_a^0=n_{q'}T$ have been employed.
By substituting Eq.~(\ref{eq:b^q}) and Eq.~(\ref{eq:g})  into Eq.~(\ref{eq:Vq'_part2}), we derive:
\begin{align}
    V^\mu_{q'}
    =&-\sum_{a}q'_a\int d\Gamma_{a}\frac{ p_a^{\langle\mu\rangle} p_a^{\langle\nu\rangle}}{\epsilon^0_a} \bigg[\sum_q\mathcal{B}_{a,\mathrm{part}}^q\nabla_\nu\alpha_q\nonumber\\
    &+\mathcal{G}_{a,\mathrm{part}}\beta  E_\nu\bigg] \bar{f}_a^0\nonumber\\
  &+\sum_{a'} \frac{n_{q'}}{\omega}\int d\Gamma_{a'}\widetilde{\epsilon}_{a'}^0\frac{p_{a'}^{\langle\mu\rangle} p_{a'}^{\langle\nu\rangle}}{\epsilon_{a'}^0}\sum_{q}\mathcal{B}_{a',\mathrm{part}}^q \bar{f}_{a'}^0\nabla_\nu \alpha_q\nonumber\\
    &+\frac{ n_{q'}}{\omega T}\sum_{a'}\int d\Gamma_{a'} \widetilde{\epsilon}_{a'}^0\frac{ p_{a'}^{\langle\mu\rangle} p_{a'}^{\langle\nu\rangle}}{\epsilon^0_{a'}} \mathcal{G}_{a',\mathrm{part}} \bar{f}_{a'}^0 E_{\nu}.
\end{align}
 Here, we emphasize that both the $\sum_{a}$ and $\sum_{a'}$ represent the summation over all the hadron species under consideration. Thus, we can combine these distinct summation indices into a single summation index:  $\sum_a A_a+\sum_{a'}B_{a'}=\sum_{a}(A_a+B_a)$. 
Accordingly, the above equation can be further simplified to:
\begin{align}\label{eq:Vq'_final}
    V^\mu_{q'}=&\sum_{a}\int d\Gamma_{a}\frac{p_a^{\langle\mu\rangle} p_a^{\langle\nu\rangle}}{\epsilon_a^0} \left[\frac{n_{q'}(\epsilon^0_a+U_a)}{\omega}-q'_a\right]\nonumber\\
&\times\left[\sum_q\mathcal{B}_{a,\mathrm{part}}^{q}\nabla_\nu\alpha_q+\beta\mathcal{G}_{a,\mathrm{part}}^{}E_\nu\right] \bar{f}_a^0.
\end{align}
We replace the right-hand side of Eq.~(\ref{eq:RTA}) with Eq.~(\ref{eq:deltaf-zeroB}), and insert $\phi_a$ from Eq.~(\ref{eq:phi_1}) into the left-hand side of Eq.~(\ref{eq:RTA}). By equating $\delta f_a $ and $\bar{f}_a^0\phi_a$ through matching tensor structure, we can derive the particular solutions for the functions $\mathcal{B}^q_{a}$ and  $\mathcal{G}_{a}$ from $\phi_a$, which are presented as follows:
\begin{eqnarray}
\mathcal{B}_{a,\mathrm{part}}^q&=&\frac{\tau_a}{\epsilon_a^0}\left[q_a-\frac{n_q}{\omega}(\epsilon^0_a+U_a)\right](1\pm \bar{f}_a^0),\\
\mathcal{G}_{a,\mathrm{part}}&=&Q_a\frac{\tau_a}{\epsilon_a^0}(1\pm \bar{f}_a^0)\label{eq:Gpart}.
\end{eqnarray}

According to the linear response theory, Eq.~(\ref{eq:Vq'_final}) can be expressed in the following matrix form: 
\begin{equation}\label{eq:kappa_matrix}
\begin{bmatrix}
V^\mu_B \\
V^\mu_Q \\
V^\mu_S\\
\end{bmatrix}=\begin{bmatrix}
\eta^{BQ}  \\
\eta^{QQ}  \\
\eta^{SQ} 
\end{bmatrix}E^{\mu}+\begin{bmatrix}
{\kappa}^{BB}&{\kappa}^{QB}&{\kappa}^{SB}\\
{\kappa}^{BQ}&{\kappa}^{QQ}&{\kappa}^{SQ}\\
{\kappa}^{BS}&{\kappa}^{QS}&{\kappa}^{SS}
\end{bmatrix}\begin{bmatrix}
\nabla^\mu\alpha_B \\
\nabla^\mu\alpha_Q \\
\nabla^\mu\alpha_S 
\end{bmatrix}.
\end{equation}
The diffusion coefficient matrix, $\kappa^{qq'}$ ($q,q'\in\{B,Q,S\}$), which quantifies the coupling between the diffusion of various conserved charges, is expressed as 
\begin{eqnarray}\label{eq:Kappa}
\kappa^{qq'}&=&\sum_a\frac{d_a}{3}\int \frac{d^3p_a}{(2\pi)^3}\tau_a\frac{\bm{p}_a^2}{(\epsilon^0_a)^2}\left[q'_a-(\epsilon_a^0+U_a)\frac{n'_q}{\omega}\right]\nonumber\\
&&\times\left[q_a-(\epsilon_a^0+U_a)\frac{n_q}{\omega}\right]\bar{f}^0_a(1\pm\bar{f}^0_a).
\end{eqnarray}
This expression is equivalent to the one presented in Ref.~\cite{Das:2021bkz}, excluding the effects of quantum statistics and the repulsive mean-field interactions.
In Eq.~(\ref{eq:kappa_matrix}), the thermoelectric transport coefficient matrix, $\eta^{qq'}$, is defined as  
\begin{eqnarray}\label{eq:eta}
\eta^{qq'}&=&\sum_a\frac{d_a\beta}{3}\int \frac{d^3p_a}{(2\pi)^3}\tau_a\frac{\bm{p}_a^2}{(\epsilon^0_a)^2}q'_a\nonumber\\
&&\times\left[q_a-(\epsilon_a^0+U_a)\frac{n_q}{\omega}\right]\bar{f}^0_a(1\pm \bar{f}^0_a).
\end{eqnarray} 
By setting $q_a'=Q_a$, we redefine $\eta^{qQ}$ as the thermoelectric conductivity associated with the conserved charge $q$.
When the  net electric diffusion current vanishes, i.e., $V^i_Q=\bm{V}_Q=0$, the induced electric field is given  by:
\begin{equation}\label{eq:inerelectric}
\bm{ E}_{}^{}=\sum_qM^{qQ}T\bm{ \nabla}\alpha_q.
\end{equation}
Here, $M^{qQ}$ is defined as diffusion thermopower associated with the conserved charge  $q$, which quantifies the ability of hadronic matter to convert the gradients in conserved charge chemical thermal potentials into an electric field, and it is expressed as:
\begin{eqnarray}\label{eq:MqQ}
M^{qQ}=\beta\kappa^{qQ}/\eta^{QQ}.
\end{eqnarray}
By inserting Eq.~(\ref{eq:inerelectric}) into Eq.~(\ref{eq:kappa_matrix}),  Eq.~(\ref{eq:Vq'_final}) can be rewritten as 
\begin{equation}\label{eq:modified_matrix}
\begin{bmatrix}
V^\mu_B \\
V^\mu_Q \\
V^\mu_S 
\end{bmatrix}=\begin{bmatrix}
\widetilde{\kappa}^{BB}~\widetilde{\kappa}^{QB}~\widetilde{\kappa}^{SB}\\
\widetilde{\kappa}^{BQ}~\widetilde{\kappa}^{QQ}~\widetilde{\kappa}^{SQ}\\
\widetilde{\kappa}^{BS}~\widetilde{\kappa}^{QS}~\widetilde{\kappa}^{SS}
\end{bmatrix}\begin{bmatrix}
\nabla^\mu\alpha_B \\
\nabla^\mu\alpha_Q \\
\nabla^\mu\alpha_S 
\end{bmatrix}.
\end{equation}
In this formulation, the thermoelectric modified diffusion coefficients in the electric current sector vanish, specfically, $\widetilde{\kappa}^{BQ}=\widetilde{\kappa}^{SQ}=\widetilde{\kappa}^{QQ}=0$. The thermoelectric modified diffusion coefficient matrix elements in the baryon and strangeness diffusion current sectors takes the  form:
\begin{eqnarray}\label{eq:kappaqq'}
\widetilde{\kappa}^{qq''}=\kappa^{qq''}-TM^{qQ}\eta^{q''Q}.
\end{eqnarray}
Here, $q''\in\{B,S\}$.

%-----------------------------------------------------------------------------------
\subsection{For finite magnetic field}
Next, we investigate the influence of the magnetic field on the thermoelectric effect and diffusion processes involving multiple conserved charges in the hadronic medium. In the presence of a weak magnetic field,  the field is not the dominant energy scale and its impact manifest primarily  at the classical level through the cyclotron motion of charged particles. We reasonably propose that the scattering mechanism of the constituents and the thermodynamic quantities remain unaffected by the magnetic field. In the uniform electric field $\bm{E}$ and external magnetic field $\bm{H}$  ($\bm{E}\perp \bm{H}\perp\bm{ \nabla}\alpha_q$),  taking into account a time-independent  phase space distribution function, Eq.~(\ref{eq:Boltzmann_eq}) can be reformulated as follows: 
\begin{equation}\label{eq:boltzmann}
\bm{v}_{a}\cdot \bm{\nabla}
f_{a} +[Q_{a}(\bm{E}^{}+\bm{v}_{a}\times\bm{H})-\bm{\nabla} U_a]\cdot\frac{\partial f_{a}}{\partial 
\bm{p}_a}=-\frac{\delta f_a}{\tau_a}.
\end{equation}
where the RTA has been employed, and  $\bm{v}_{a}\equiv\frac{d\epsilon^0_{a}}{d\bm{p}_a}=\bm{p}_a/\epsilon^0_{a}$ is the three-velocity of particle species $a$. Neglecting any second-order terms in $\delta f_a$, Eq.~(\ref{eq:boltzmann}) can be simplified as 
\begin{align}\label{eq:boltzmann2}
 \frac{f_a}{\tau_{a}}+Q_{a}\bm{ v}_a\times\bm{ H}\cdot\frac{\partial f_a}{\partial\bm{p}_a}=&-Q_{a}\bm{E}\cdot\frac{\partial \bar{f}_a^0}{\partial\bm{p}_a}-(\bm{v}_{a}\cdot
\bm{\nabla}\bar{f}^0_a)\nonumber\\
 &+\frac{\bar{f}_a^0}{\tau_{a}}+\bm{\nabla} U_a\cdot\frac{\partial \bar{f}_a^0 }{\partial\bm{p}_a}.
\end{align}
We further assume  the solution to Eq.~(\ref{eq:boltzmann2}) 
satisfies the following linear form:
\begin{align}\label{eq:fa}
f_{a}=&\bar{f}_{a}^{0}-\tau_{a}Q_a\bm{E}\cdot\frac{\partial 
\bar{f}_{a}^{0}}{\partial\bm{p}_a}-\bm{\Xi}\cdot\frac{\partial 
\bar{f}_{a}^{0}}{\partial \bm{p}_a}-\tau_{a}\bm{v}_a\cdot
\bm{\nabla}\bar{f}_{a}^{0}\nonumber\\
&+\tau_a\bm{\nabla} U_a\cdot\frac{\partial \bar{f}_a^0 }{\partial \bm{p}_a},
\end{align}
where $\bm{\Xi}$ is an unknown quantity related to the magnetic field. 
By inserting Eq.~(\ref{eq:fa}) into Eq.~(\ref{eq:boltzmann2}), we derive:
\begin{align}
 &-\frac{\bm{\Xi}}{\tau_a}\cdot\frac{\partial 
\bar{f}_{a}^{0}}{\partial \bm{p}_a}+  Q_a\bm{v}_a\times \bm{H}\cdot \frac{\partial }{\partial \bm{p}_a}\bigg[\bar{f}_a-\tau_{a}Q_a\bm{E}\cdot\frac{\partial 
\bar{f}_{a}^{0}}{\partial\bm{p}_a}\nonumber\\
&-\bm{\Xi}\cdot\frac{\partial 
\bar{f}_{a}^{0}}{\partial \bm{p}_a}-\tau_{a}\bm{v}_a\cdot
\bm{\nabla}\bar{f}_{a}^{0}+\tau_a\bm{\nabla} U_a\cdot\frac{\partial \bar{f}_a^0 }{\partial \bm{p}_a}\bigg]=0.\label{eq:f_a3}
\end{align}
 To facilitate the calculation, we provide the following identities,
\begin{align}
     \frac{\partial \bar{f}_a^0}{\partial \bm{p}_a}=&-\bar{f}_a^0(1\pm \bar{f}_a^0)\beta\bm{v}_a,\\
        \frac{\partial \bar{f}_a^0(1\pm \bar{f}_a^0)}{\partial \bm{p}_a}=&-(1\pm 2\bar{f}_a^0)\bar{f}_a^0(1\pm\bar{f}_a^0)\beta\bm{v}_a,\\
    \frac{\partial (\epsilon_a^0)^{-1}}{\partial \bm{p}_a}=&-\frac{\bm{v}_a}{(\epsilon_a^0)^2}.
\end{align}
 Then, several terms in Eq.~(\ref{eq:f_a3}) can be calculated  as 
\begin{align}
 \frac{\partial}{\partial \bm{p}_a}\left[\bm{E}\cdot\frac{\partial\bar{f}_a^0}{\partial\bm{p}_a}\right]=&\bigg[(\bm{E}\cdot \bm{v}_a)(1\pm 2\bar{f}_a^0)\bar{f}_a^0(1\pm \bar{f}_a^0)\bm{v}_a\beta^2\nonumber\\
   &-\bm{E} \bar{f}_a^0(1\pm \bar{f}_a^0)\frac{\beta}{\epsilon_a^0}\nonumber\\
    &+(\bm{E}\cdot \bm{v}_a) \bar{f}_a^0(1\pm \bar{f}_a^0)\frac{\bm{v}_a}{\epsilon_a^0}\beta\bigg],\label{eq:cal_part1}\\
      \frac{\partial}{\partial \bm{p}_a}\left[\bm{\Xi}\cdot\frac{\partial\bar{f}_a^0}{\partial\bm{p}_a}\right]=&\bigg[(\bm{\Xi}\cdot \bm{v}_a)(1\pm 2\bar{f}_a^0)\bar{f}_a^0(1\pm \bar{f}_a^0)\bm{v}_a\beta^2\nonumber\\
       &-\bm{\Xi} \bar{f}_a^0(1\pm \bar{f}_a^0)\frac{\beta}{\epsilon_a^0}\nonumber\\
    &+(\bm{\Xi}\cdot \bm{v}_a) \bar{f}_a^0(1\pm \bar{f}_a^0)\frac{\bm{v}_a}{\epsilon_a^0}\beta\bigg],\label{eq:cal_part2}\\
\frac{\partial }{\partial \bm{p}_a}\left[\bm{v}_a\cdot\bm{\nabla}\bar{f}_{a}^{0}\right]=&
\bigg[\sum_q\big(\frac{n_q}{\omega}(\epsilon_a^0+U_a)-q_a\big)\bm{v}_a\cdot\bm{\nabla}\alpha_q\nonumber\\
&+\beta\bm{v}_a\cdot\bm{\nabla}U_a\bigg]\bar{f}_a^0(1\pm \bar{f}_a^0)\nonumber\\
&\times\bigg[(1\pm 2\bar{f}_a^0)\beta\bm{v}_a-\frac{1 }{\epsilon_a^0}+\frac{ \bm{v}_a}{\epsilon_a^0}\bigg]\nonumber\\
   &-(\bm{v}_a\cdot\bm{\nabla}\beta) \bm{v}_a \bar{f}_a^0(1\pm \bar{f}_a^0),\label{eq:cal_part3}\\
   \frac{\partial }{\partial \bm{p}_a}\left[\bm{\nabla}U_a \frac{\partial \bar{f}_a^0}{\partial \bm{p}_a}\right]=&-\bm{\nabla}U_a \frac{\beta }{\epsilon_a^0}\bar{f}_a^0(1\pm\bar{f}_a^0)\nonumber\\
  & +(\bm{\nabla}U_a\cdot \bm{v}_a) \frac{\beta}{\epsilon_a^0} \bar{f}_a^0(1\pm\bar{f}_a^0)\bm{v}_a\nonumber\\
   &+(\bm{\nabla}U_a\cdot \bm{v}_a)\beta^2 (1\pm 2\bar{f}_a^0)\bar{f}_a^0(1\pm\bar{f}_a^0)\bm{v}_a.\label{eq:cal_part4}
\end{align}
 Since Eqs.~(\ref{eq:cal_part1}-\ref{eq:cal_part4})  involve a dot product with $\bm{v}_a\times \bm{H}$, certain vector components within these equations become irrelevant by {\tiny }the orthogonality condition  $(\bm{v}_a\times \bm{H})\cdot\bm{v}_a$=0. Consequently, Eq.~(\ref{eq:f_a3}) can be simplified and rewritten as
\begin{align}\label{eq:eq1}
0=&\frac{F_a}{\tau_a}\bm{\Xi}\cdot\bm{v}_a+\frac{Q_a}{\epsilon_a^0}F_a\bm{v}_a\times \bm{H}\cdot\bm{\Xi}+\frac{Q^2_a}{\epsilon_a^0}\tau_aF_a\bm{v}_a\times \bm{H}\cdot\bm{E}\nonumber\\
&+\frac{Q_a\tau_a}{\epsilon_a^0}\bm{v}_a\times\bm{H}\cdot\sum_{q}H_a^q\bm{\nabla}\alpha_q.
\end{align} 
Through rigorous calculations, we obtain the following expression for $\bm{\Xi}$: \begin{align}\label{eq:Xi}
\bm{\Xi}=&\frac{\tau_a}{(\omega_{c,a}\tau_a)^2+1}[-\omega_{c,a}^2Q_a\bm{E}+\frac{\omega_{c,a}}{\tau_a}Q_a(\bm{E}\times \bm{h})
\nonumber\\
&+\frac{\omega_{c,a} H_a^q}{\tau_aF_a}(\bm{\nabla}\alpha_q\times \bm{h})-\frac{\omega_{c,a}^2H_a^q}{F_a}\bm{\nabla}\alpha_q],
\end{align}
where $\bm{h}=\bm{H}/H$. The magnetic field information is embedded in cyclotron frequency of particle species $a$, denoted by $\omega_{c,a}^{}=Q_{a}H/\epsilon^0_{a}$. By  inserting Eq.~(\ref{eq:Xi}) into  Eq.~(\ref{eq:fa}), we finally obtain the  magnetic field-dependent perturbation term of the distribution function,
\begin{align}\label{eq:deltaf}
\delta f^{}_a=&\frac{\tau_a^2}{(\omega_{c,a}\tau_a)^{2}+1}\big[\omega_{c,a}F_aQ_a(\bm{E}\times \bm{h})\cdot \bm{v}_a\nonumber\\
&+\frac{1}{\tau_a}F_aQ_a\bm{E}\cdot\bm{v}_a+\sum_q\omega_{c,a}H^q_a(\bm{\nabla}\alpha_q\times \bm{h})\cdot\bm{v}_a\nonumber\\
&+\sum_q\frac{1}{\tau_a}H^q_a\bm{v}_a\cdot\bm{\nabla}\alpha_q\big].
\end{align}
Repeating some procedures in the previous subsection (Eqs.~(\ref{f_phi})-(\ref{eq:deltaV}), Eqs.~(\ref{eq:LLcondition})-(\ref{eq:Gpart})), the particular  solutions for the functions $\mathcal{B}^q_{a}$ and $\mathcal{G}_{a}$ from $\phi_a$  in the presence of a magnetic field are computed as 
\begin{align}
\mathcal{B}_{a,\mathrm{part}}^{q,H\neq0}=&\frac{\tau_a}{\epsilon_a^0}\left[q_a-\frac{n_q}{\omega}(\epsilon^0_a+U_a)\right](1\pm \bar{f}_a^0)\nonumber\\
&\times\frac{[\omega_{c,a}\tau_a(\bm{h}\times\bm{v}_a)+1]}{(\omega_{c,a}\tau_a)^2+1},\label{eq:B_part_H}\\
\mathcal{G}_{a,\mathrm{part}}^{H\neq0}=&\frac{Q_a\tau_a}{\epsilon_a^0}(1\pm \bar{f}_a^0)\frac{[1+\omega_{c,a}\tau_{a}(\bm{h}\times\bm{v}_a)]}{(\omega_{c,a}\tau_a)^2+1}\label{eq:G_part_H}.
\end{align}
Upon substitution of Eq.~(\ref{eq:B_part_H}) and Eq.~(\ref{eq:G_part_H}) into Eq.~(\ref{eq:Vq'_final}), the diffusion current of the conserved charge $q'$ in a magnetic field is given by:
\begin{align}
\bm{V}^{q'}=&-\sum_{a,q}\int d\Gamma_{a}\frac{\bm{p}_a}{(\epsilon^0_a)^2}\frac{\tau_a\bm{p}_a\cdot[\omega_{c,a}\tau_a(\bm{\nabla}\alpha_q\times\bm{h})+\bm{\nabla}\alpha_{q}]}{(\omega_{c,a}\tau_a)^2+1}\nonumber\\
&\times\left[q_a-(\epsilon_a^0+U_a)\frac{n_{q}}{\omega}\right]\left[q'_a-(\epsilon_a^0+U_a)\frac{n_{q'}}{\omega}
\right]\nonumber\\
&\times \bar{f}_a^0(1\pm \bar{f}_a^0)\nonumber\\
&+\sum_{a} q'_a\int d\Gamma_{a}\frac{\bm{p}_a}{(\epsilon^0_a)^2} \frac{Q_a}{T}\frac{\tau_a\bm{p}_a\cdot\left[\omega_{c,a}\tau_a(\bm{E}\times\bm{h})+\bm{E}\right]}{(\omega_{c,a}\tau_a)^2+1}\nonumber\\
&\times\left[q'_a-\frac{n_{q'}}{\omega}(\epsilon^0_a+U_a)\right]\bar{f}_a^0(1\pm \bar{f}_a^0).
\end{align}
Assuming  the magnetic field is aligned with the $z$-axis,  the above equation can be decomposed into $x$- and $y$-components, resulting in the following matrix form:
\begin{equation}\label{eq:Vmatrix1}
\begin{bmatrix}
V^{q'}_x \\
V^{q'}_y 
\end{bmatrix}=	\begin{bmatrix}
\eta^{Qq'Q}_{xx}~\eta^{Qq'Q}_{xy} \\
\eta^{Qq'Q}_{yx}~\eta^{Qq'Q}_{yy} 
\end{bmatrix}\begin{bmatrix}
E_{x} \\
E_{y} 
\end{bmatrix}+\sum_{q}\begin{bmatrix}
{\kappa}_{xx}^{Qqq'}~{\kappa}_{xy}^{Qqq'} \\
{\kappa}_{yx}^{Qqq'}~{\kappa}_{yy}^{Qqq'} 
\end{bmatrix}\begin{bmatrix}
-\nabla_x\alpha_{q} \\
-\nabla_y\alpha_{q} 
\end{bmatrix}.
\end{equation}
Here, the thermoelectric conductivity tensors,  denoted as $\eta^{Qq'Q}$, and diffusion coefficient tensors, denoted as  $\kappa^{Qqq'}$, in a magnetic field satisfy the Onsager's reciprocity relation~\cite{Callen,Redin}:  $\kappa^{Qqq'}_{xx}(\eta^{Qq'Q}_{xx})=\kappa_{yy}^{Qqq'}(\eta_{yy}^{Qq'Q})$ and $\kappa^{Qqq'}_{xy}(\eta^{Qq'Q}_{xy})=-\kappa_{yx}^{Qqq'}(-\eta_{yx}^{Qq'Q})$.
Accordingly, the magnetic field-dependent thermoelectric conductivity matrix element, $\eta_{xx}^{Qq'Q}$, and the Hall-like or transverse thermoelectric conductivity matrix element, $ \eta_{yx}^{Qq'Q}$, can be expressed as
\begin{eqnarray}\label{eq:eta-xx-xy}
\begin{bmatrix}
\eta_{xx}^{Qq'Q}\\
\eta_{yx}^{Qq'Q}
\end{bmatrix}=&&\sum_{a}\frac{d_aQ_{a}}{3T}\int\frac{d^3p_a}{(2\pi)^3}\frac{\bm{p}_a^2}{(\epsilon_{a}^0)^2}\frac{\left(q'_a-(\epsilon_a^0+U_a)\frac{n_{q'}}{\omega}\right)}{(\omega_{c,a}\tau_a)^2+1}\nonumber\\
&&\times\begin{bmatrix}
\tau_{a}\\
-\omega_{c,a}\tau_a^2
\end{bmatrix}	\bar{f}_a^{0}(1\pm\bar{f}_a^{0}).
\end{eqnarray} 
Similarly, the magnetic field-dependent diffusion coefficient matrix element, $\kappa_{xx}^{Qqq'}$, and the Hall-like diffusion coefficient matrix element, $\kappa_{yx}^{Qqq'}$, can be given as
\begin{align}\label{eq:kappa-xx-xy}
\begin{bmatrix}
\kappa_{xx}^{Qqq'}\\
\kappa_{yx}^{Qqq'}
\end{bmatrix}=&\sum_{a}\frac{d_a}{3}\int\frac{d^3p_a}{(2\pi)^3}\frac{\bm{p}_a^2}{(\epsilon_{a}^0)^2}\frac{\left(q_a-(\epsilon_a^0+U_a)\frac{n_q}{\omega}\right)}{(\omega_{c,a}\tau_a)^2+1}\nonumber\\
&\times\left(q'_a-(\epsilon_a^0+U_a)\frac{n_{q'}}{\omega}\right)	\begin{bmatrix}
\tau_{a}\\
-\omega_{c,a}\tau_a^2
\end{bmatrix}\nonumber\\
&\times\bar{f}_a^{0}(1\pm\bar{f}_a^{0}).
\end{align} 
To better distinguish the expressions of transport coefficients under various transverse restriction conditions later, we convert Eq.~(\ref{eq:Vmatrix1}) into the following matrix form:
\begin{widetext}
\begin{align}  \label{eq:Vmatrix2}
\begin{bmatrix}  
E^{}_x \\
E^{}_y \\
V^{q''}_x\\
V^{q''}_y
\end{bmatrix} =   
\begin{bmatrix}  
\rho^{QQ}_{xx} & \rho^{QQ}_{xy}&\sum_{q}-TM^{QqQ}_{xx}&\sum_{q}-TM^{QqQ}_{xy}\\
\rho^{QQ}_{yx}&\rho^{QQ}_{yy}&\sum_{q}-TM^{QqQ}_{yx}&\sum_{q}-TM^{QqQ}_{yy}\\
\Pi_{xx}^{Qq''Q}&\Pi_{xy}^{Qq''Q}&\sum_{q}\widetilde{\kappa}_{xx}^{Qqq''}&\sum_{q}\widetilde{\kappa}_{xy}^{Qqq''}\\
\Pi_{yx}^{Qq''Q}&\Pi_{yy}^{Qq''Q}&\sum_{q}\widetilde{\kappa}_{yx}^{Qqq''}&\sum_{q}\widetilde{\kappa}_{yy}^{Qqq''}
\end{bmatrix}  \begin{bmatrix}  
V^Q_{x} \\
V^Q_{y} \\
-{\nabla}_x\alpha_{q}\\
-{\nabla}_y\alpha_{q}
\end{bmatrix},~~~q\in\{B,Q,S\},~q''\in\{B,S\}.
\end{align}
\end{widetext}
Solving  the set of coupled matrix equations~(\ref{eq:Vmatrix2}) under the condition of vanishing gradients of transverse  conserved charge chemical thermal potentials ($\nabla_y\alpha_q=0$) with  $\nabla_x\alpha_q=0=V_y^Q=0$, then the electric  resistance tensors, $\rho^{QQ}$,s can be obtained as 
\begin{eqnarray}
\rho_{xx}^{QQ}=&\rho_{yy}^{QQ}=\frac{E_x}{V_x^Q}\bigg|_{\nabla_y\alpha_q=0}=\frac{\eta_{xx}^{QQQ}}{(\eta_{xx}^{QQQ})^2+(\eta_{xy}^{QQQ})^2},\\
\rho_{yx}^{QQ}=&-\rho_{xy}^{QQ}=\frac{E_y}{V_x^Q}\bigg|_{\nabla_y\alpha_q=0}=\frac{\eta_{xy}^{QQQ}}{(\eta_{xx}^{QQQ})^2+(\eta_{xy}^{QQQ})^2}.
\end{eqnarray}
In a magnetic field, the Hall-like diffusion thermopower of conserved charge $q$, denoted as  $M_{yx}^{QqQ}$, can emerge. Under the condition of $\nabla_y\alpha_q=0$ with $V_x^Q=V_y^Q=0$, the magnetic field-dependent diffusion thermopower and Hall-like diffusion thermopower of conserved charge $q$ can  be derived respectively as
\begin{align}
M_{xx}^{QqQ}&=\frac{{ E}_x}{T{\nabla}_x\alpha_{q}}\bigg|_{\nabla_y\alpha_q=0}
=\beta\rho_{xx}^{QQ}\kappa_{xx}^{QqQ}-\beta\rho_{xy}^{QQ}\kappa_{xy}^{QqQ},\\
M_{yx}^{QqQ}&=\frac{{ E}_y}{T{\nabla}_x\alpha_{q}}\bigg|_{\nabla_y\alpha_q=0}=
-\beta\rho_{xx}^{QQ}\kappa_{xy}^{QqQ}-\beta\rho_{xy}^{QQ}\kappa_{xx}^{QqQ}.
\end{align}
In the absence of a magnetic field,  $M_{xx}^{QqQ}$ simplifies to $M^{Qq}$.
Utilizing the set of coupled equations~(\ref{eq:Vmatrix2}), and subjecting to the condition of $\nabla_y\alpha_q=0$ with $V_x^Q=V_y^Q=0$, the magneto-thermoelectric modified diffusion coefficient, $\widetilde{\kappa}_{xx}^{Qqq''}$, and Hall-like magneto-thermoelectric modified  diffusion coefficient, $\widetilde{\kappa}_{yx}^{Qqq''}$,  can  be computed respectively as 
\begin{align}
\widetilde{\kappa}_{xx}^{Qqq''}&=\frac{{ V}_x^{q''}}{-{\nabla}_x\alpha_{q}}\bigg|_{\nabla_y\alpha_q=0}\nonumber\\&={\kappa}_{xx}^{Qqq''}-\eta^{Qq''Q}_{xx}TM_{xx}^{QqQ}+\eta^{Qq''Q}_{yx}TM_{yx}^{QqQ},\\
\widetilde{\kappa}_{yx}^{Qqq''}&=\frac{{ V}_y^{q''}}{-{\nabla}_x\alpha_{q}}\bigg|_{\nabla_y\alpha_q=0}\nonumber\\&={\kappa}_{yx}^{Qqq''}-\eta^{Qq''Q}_{yx}TM_{xx}^{QqQ}-\eta^{Qq''Q}_{xx}TM_{yx}^{QqQ}.
\end{align}

In condensed matter physics, the Righi-Leduc effect or thermal Hall effect occurs when a transverse temperature gradient ($\nabla_yT$) is developed by a longitudinal temperature gradient ($\nabla_xT$) under a static magnetic field ($H_z$). The associated Righi-Leduc coefficient is determined  under the transverse adiabatic condition, specifically when the transverse heat current ($I_y$) is zero. Similarly, in hadronic matter with multiple conserved charges,  a transverse or Hall-like conserved charge density gradient ($\nabla_{y}\alpha_q$) perpendicular to both the $\nabla_{x}\alpha_q$ and $H_z$, can be induced. The corresponding coefficient is calculated under the condition of vanishing transverse diffusion current i.e., $V_y^{q''}=0$.
Therefore, utilizing the matrix Eq.~(\ref{eq:Vmatrix2}), and enforcing the conditions  $V_{y}^{q''}=V_x^Q=V_y^Q=0$, the Righi-Leduc-type relation within the hadronic medium can be deduced as follows: 
\begin{align}
0=&\sum_q\widetilde{\kappa}_{yx}^{Qqq''}\nabla_x\alpha_q+\sum_{q}\widetilde{\kappa}_{xx}^{Qqq''}\nonumber\nabla_{y}\alpha_q\\
\to& \mathcal{L}^{q}_{V_{y}^{q''}=0}=\frac{\nabla_y\alpha_{q}}{\nabla_x\alpha_{q}}\bigg|_{V_{y}^{q''}=0}  =\frac{\widetilde{\kappa}_{xy}^{Qqq''}}{\widetilde{\kappa}_{xx}^{Qqq''}},
\end{align} 
where $\mathcal{L}^{q}_{V_{y}^{q''}=0}$ denotes Righi-Leduc-like coefficient.
Accordingly, the magnetic field-dependent diffusion thermopower  
under the condition of $V_{y}^{q''}=0$ 
with $V_x^Q=V_y^Q=0$ is computed as  
\begin{eqnarray}
M^{QqQ}_{xx,{V_y^{q''}}=0} 
&=&M_{xx}^{QqQ}-M_{yx}^{QqQ}\mathcal{L}^q_{V_{y}^{q''}=0}.\label{eq:Adia_Mxx}
\end{eqnarray}
When the magnetic field is turned off, the expressions of transport coefficients become equivalent  under various transverse condition: $\nabla_y\alpha_{q}=0$ and $V_y^{q''}=0$. Similarly, the magneto-thermoelectric modified diffusion coefficient
under the condition of $V^{q''}_y=0$ with $V_x^Q=V_y^Q=0$, is computed as
\begin{eqnarray}
\widetilde{\kappa}_{xx,V_y^{q''}=0}^{Qqq''}&=&\widetilde{\kappa}_{xx}^{Qqq''}-\widetilde{\kappa}_{yx}^{Qqq''}\mathcal{L}^{q}_{V_y^{q''}=0}\label{eq:ad_X_Kappa}.
\end{eqnarray}

\subsection{Thermal averaged relaxation time}\label{sec:tau}
Comparing Eq.~(\ref{eq:solution}) and Eq.~(\ref{eq:RTA}),  the mutual interaction information of all particle species is encoded in the relaxation time, $\tau_a$.  For the binary elastic collisions $a (p_a)+b(p_b')\to a(p_a'')+b(p_b''')$, the inverse of $\tau_a$ is given as 
\begin{align}
\tau_a^{-1}=&\sum_{b}\gamma_{ab}\int d\Gamma_{b}'d\Gamma_{a}''d\Gamma_{b}'''W_{ab}(p_a,p_b'|p_a'',p_b''')
\nonumber\\
&\times\bar{f}_b^{0'}(1\pm \bar{f}_{a}^{0''}) \frac{(1\pm \bar{f}_b^{0'''})}{1\pm \bar{f}_a^0},
\end{align}
where the transition rate is given as 
\begin{align}\label{eq:M}
 W_{ab}
    =\frac{(2\pi)^4\delta^4(p_a+p_b'-p_a''-p_b''')}{16\epsilon_a^0\epsilon_b^{0'}\epsilon_a^{0''}\epsilon_d^{0''}}|\bar{\mathcal{M}}_{ab\to ab}|^2.
\end{align}
Here, $|\bar{\mathcal{M}}|$ is the dimensionless transition amplitude averaged over the spin degeneracy factor in both initial and final states~\cite{Larionov:2007hy}. This is necessary to balance
 the degeneracy factors in the $d\Gamma_a$.
To simplify the estimation of relaxation time, $(1\pm \bar{f}_a^{0''}) \frac{(1\pm \bar{f}_b^{0'''})}{1\pm \bar{f}_a^0}\simeq1$, and we utilize the formula of scattering cross section~\cite{Peskin:1995ev}
\begin{align}
\sigma_{ab}=\frac{\int d\Gamma_{a}''d\Gamma_{b}'''(2\pi)^{4}\delta^{4}(p_a+p_b'-p_a''-p_b''')|\bar{\mathcal{M}}|^{2}}{16\epsilon_a^{0''}\epsilon_b^{0'''}\sqrt{(p_{a}\cdot p_{b}')^{2}-m_{a}^{2}m_{b}^{2}}},
\end{align}
then we can rewrite $\tau_{a}^{-1}$ as 
\begin{eqnarray}\label{eq:tau_a}
\tau_{a}^{-1}=\sum_{b}^{}\gamma_{ab}\rho_{b}\sigma_{ab} 
v_{ab},
\end{eqnarray}
where$\rho_{b}=\int d\Gamma_{b}' \bar{f}_{b}^{0'}$ is the number density
of hadron species $b$. It is worth noting that the RMF interactions between hadrons can influence the scattering process by modifying the number density of hadron species $b$. Therefore, the number density of hadron species $b$ needs to be distinguished in the IHRG model and the  RMFHRG model. In Eq.~(\ref{eq:tau_a}), the relative velocity $v_{ab}$ is defined as
\begin{align}
v_{ab}=\frac{\sqrt{(\epsilon^0_{a}\epsilon_{b}^{0'}-\bm{p}_{a} \cdot	\bm{p}_{b}')^{2}-m_{a}^{2}m_{b}^{2}}}{\epsilon^0_{a}\epsilon^{0'}_{b}}.
\end{align}
We shall consider the momentum-independent relaxation time, the thermal averaged cross section $\langle\sigma_{ab} 
v_{ab}\rangle$ can be given as 
\begin{eqnarray}
\langle\sigma_{ab}  v_{ab}\rangle=\frac{\int 	d^{3}p_{a}d^{3}p_{b}'\bar{f}_{a}^{0}\bar{f}_{b}^{0'}\sigma_{ab}v_{ab}}{\int d^{3}p_{a}d^{3}p_{b}'\bar{f}_{a}^{0}\bar{f}_{b}^{0'}}.
\end{eqnarray}
In this study,  all hadrons are regarded as hard spheres with the same radius $r_{h}$, and  $\sigma_{ab}$ is a constant given by $\sigma_{ab}=4\pi r_{h}^2=30~$mb
%-----------------------------------------------------------------------------------
\begin{figure*}[htpb]
\centering    
\subfloat{\includegraphics[scale=0.46]{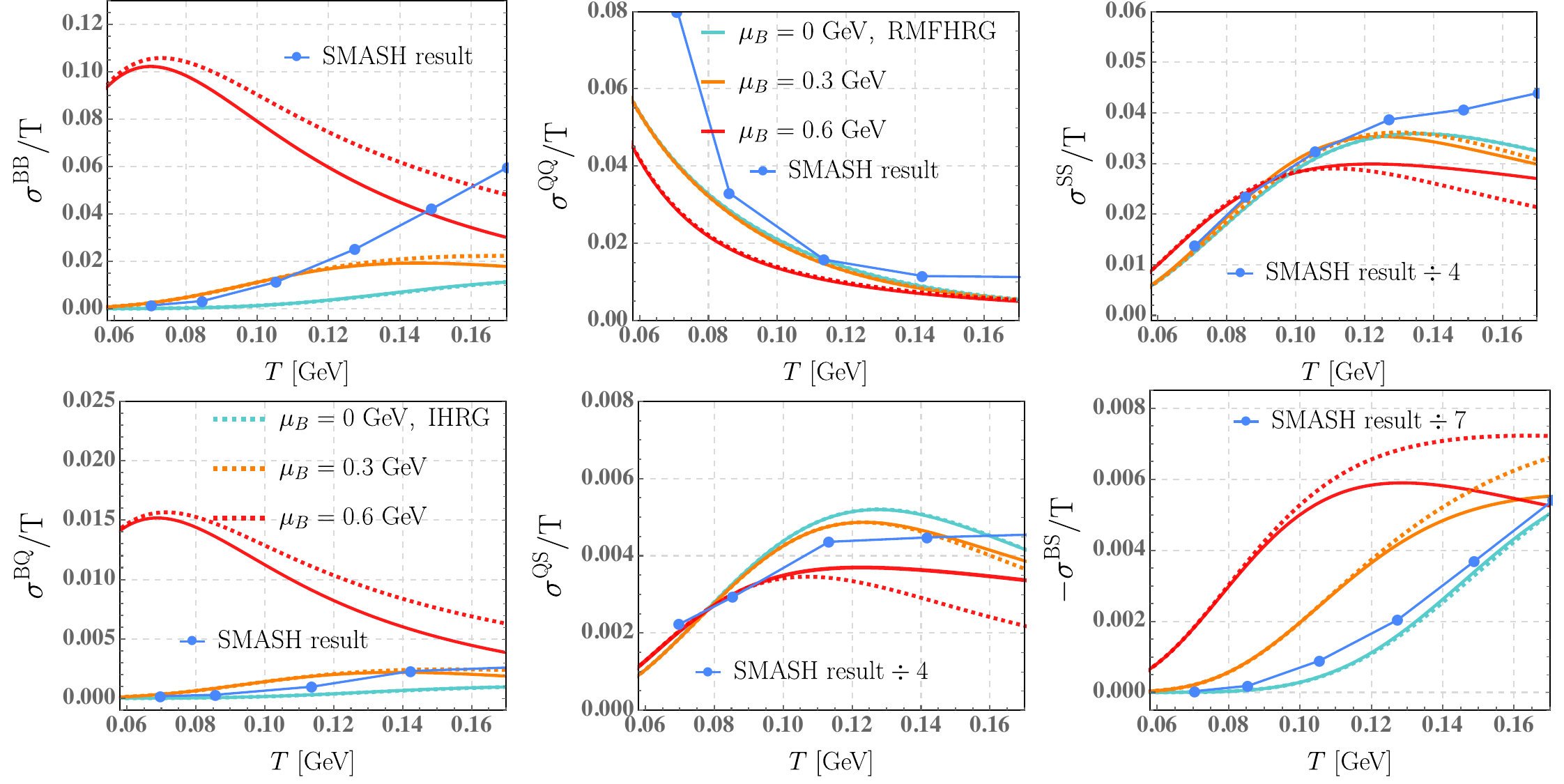}} 
\caption{The temperature dependence of complete scaled conductivity matrix $\sigma^{qq'}/T$ for different baryon chemical potentials, i.e., $\mu_{B}=0~$ (blue),~0.3~GeV (orange) and 0.6~GeV (red) in the IHRG model (dashed lines) and RMFHRG model (solid lines). 
The symbol lines are the estimations in the full SMASH simulation using the Green-Kubo formalism at zero $\mu_{B}$~\cite{Hammelmann:2023fqw}.} 
\label{fig_electric_conductivity}
\end{figure*}
\section{numerical results and discussions}\label{sec:results}
All calculations are performed in the condition of $n_S=0=\mu_{Q}=0$~\cite{Fotakis:2019nbq, Das:2021bkz}, which is expected in the initial stages of heavy-ion collision~\cite{Cleymans:1992zc,Greiner:1987tg}. This specific condition gives rise to a nonzero strangeness chemical potential, which is a function of $T$ and $\mu_B$. In the HRG models, we include all hadrons listed in  \texttt{Thermal-FIST} package,  with a mass cutoff set at $\Lambda=3.0$ GeV~\cite{Vovchenko:2019pjl}. 
The BES program at RHIC covers beam energy from $\sqrt{s_{NN}}=3$~GeV to 200~GeV, with the baryon chemical potential ranging from $\mu_{B}\simeq 0.75$~GeV to 0.02~GeV~\cite{Cleymans:2005xv,Odyniec:2013kna,STAR:2022vlo}. In this investigation, we focus on values of  $\mu_{B}$ up to 0.6~GeV. 
\subsection{Results for vanishing magnetic field}
To better understand the behaviors of the diffusion coefficient matrix later, we first thoroughly discuss the $T$ and $\mu_{B}$ dependence of the scaled conductivity matrix $\sigma^{qq'}/T$, which is given as 
\begin{eqnarray}\label{eq:sigma}
\frac{\sigma^{qq'}}{T}=\sum_a\frac{d_a\beta^2}{3}\int \frac{d^3p_a}{(2\pi)^3}\tau_a\frac{\bm{p}_a^2}{(\epsilon^0_a)^2}q'_aq_a\bar{f}^0_a(1\pm\bar{f}^0_a).
\end{eqnarray}
It is closely related to the diffusion coefficient matrix. At $\mu_{B}=0$, $\sigma^{qq'}/T$ is equal to $\kappa^{qq'}/T^2$. The variation of $\sigma^{qq'}/T$ with respect to $T$ and $\mu_{B}$ is determined by the interplay between charge number density (or distribution function) and scattering rate (or the relaxation time $\tau_{a}\sim 1/\Gamma_{\mathrm{scatt}}$). 
In Fig.~\ref{fig_electric_conductivity}, we see that the magnitude of $\sigma^{QQ}/T$ ($\sigma^{qB}/T$) for $\mu_{B}=0$ decreases (increases) monotonically as $T$ increases, on the other hand, both $\sigma^{QS}/T$ and $\sigma^{SS}/T$ first increase with $T$ and then decrease, which are qualitatively consistent with the results from the SMASH simulation  (symbol lines)~\cite{Hammelmann:2023fqw}. 
The negative sign of $\sigma^{SB}/T$ is attributed to the associated dominant carriers, i.e.,  hyperons carrying a positive baryon number with a negative strangeness. 
As shown in Fig.~\ref{fig_electric_conductivity}, all scaled conductivities obtained from the SMASH simulation using the Green-Kubo formalism are quantitively larger than ours.
We can also see that the effect of RMF interaction on the  scaled conductivity matrix for $\mu_{B}=0$ is minimal. This is because the RMF correction for $\mu_{B}=0$ results in a small suppression of charge number density and a slight enhancement of relaxation time, the mutual compensation makes the RMF correction on the conductivities negligible.

Compared to the thermal behaviors observed at $\mu_{B}=0$~GeV, the scaled conductivities at $\mu_{B}=0.3$~GeV remain largely unchanged. However, as illustrated in Fig.~\ref{fig_electric_conductivity}, there is a notable enhancement in the magnitude of $\sigma^{qB}/T$. 
This is because the dependence of $\sigma^{qB}/T$ on $\mu_{B}$ is mainly governed by the baryon density, which is an increasing function of $\mu_{B}$. 
Meanwhile, we observe a nominal reduction in both $\sigma^{QS}/T$ and $\sigma^{SS}/T$ when comparing their values at $\mu_{B}=0.3$~GeV to those at $\mu_{B}=0$ GeV. This reduction can be explained by the fact that the primary carriers for both $\sigma^{QS}/T$ and $\sigma^{SS}/T$ are kaons ($K$), the number density of kaons undergo a slight enhancement due to the nonzero $\mu_S$, this enhancement is negated by the increased scattering rates of kaons resulting from their elevated collisions with baryons. The variation in  $\sigma^{QQ}/T$  at $\mu_{B}=0.3$~GeV  is invisible compared to that at $\mu_{B}=0$~GeV.  This minimal change is the result of a competition between meson (primarily pions) and baryon (primarily nucleons) contributions to $\sigma^{QQ}$. At $\mu_{B}=0.3$~GeV, the decrease of pion contribution to $\sigma^{QQ}$ with $\mu_B$ is nearly compensated by the increase of nucleon contribution to $\sigma^{QQ}$  with $\mu_B$. 
As $\mu_{B}$ increases further, the baryon density becomes more significant, resulting in pronounced changes in all conductivities at $\mu_{B}=0.6$~GeV.
We note that in comparison to the temperature dependence at $\mu_B=0$ and 0.3 GeV, the trend of both $\sigma^{QB}/T$ and $\sigma^{BB}/T$ at $\mu_{B}=0.6$~GeV can undergo a reversal. This reversal arises because the decreasing behavior of the relaxation time for predominant nucleons with $T$ dominates over the increasing trend of the distribution function with $T$. Compared to the variation in conductivity resulting from the RMF correction at lower $\mu_B$ ($\mu_B=0$ or 0.3~GeV),  all conductivities except  $\sigma^{QQ}/T$,  undergo notable changes when the RMF correction is considered at $\mu_B=0.6$~GeV.  In particular, $\sigma^{BQ}/T$, $\sigma^{BB}/T$, and $-\sigma^{BS}/T$ at $\mu_B=0.6$~GeV show visible suppression due to the substantial reduction in baryon density of the system caused by the incorporation of the RMF correction.
%---------------------------
 \begin{figure*}[htpb]
\centering
\subfloat{\includegraphics[scale=0.43]
 {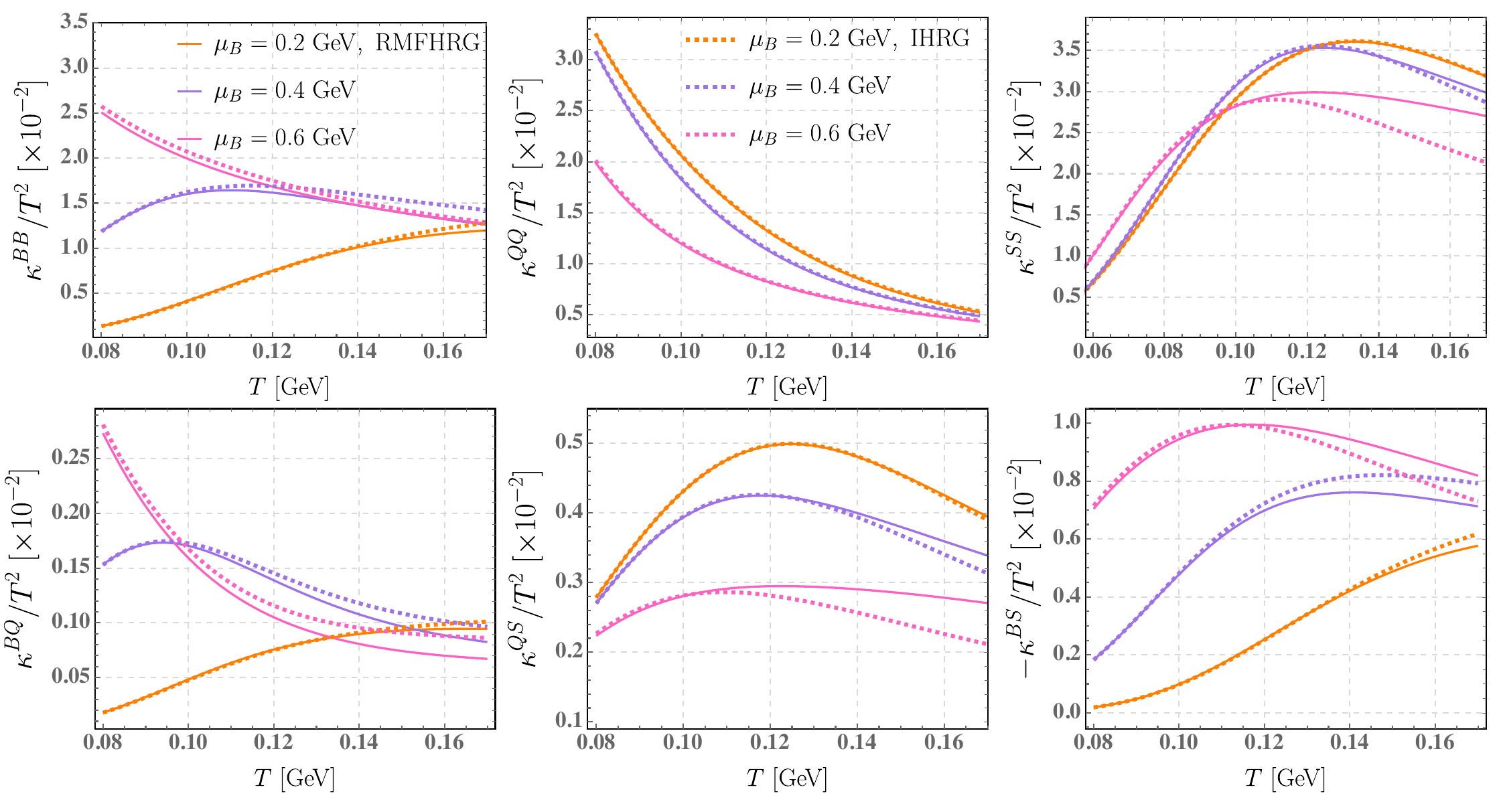}}
\caption{The temperature dependence of scaled diffusion coefficient matrix  $\kappa^{qq'}/T^2$ at $\mu_{B}=0.2$ GeV (orange),~0.4 GeV (purple),~0.6 GeV (magenta) in the IHRG model (dashed lines) and RMFHRG model (solid lines).}
\label{fig_Kappa}
\end{figure*} 
%-----------------------------------------------------------------------------------
\begin{figure*}[htpb]
\centering
\subfloat{\includegraphics[scale=0.43]{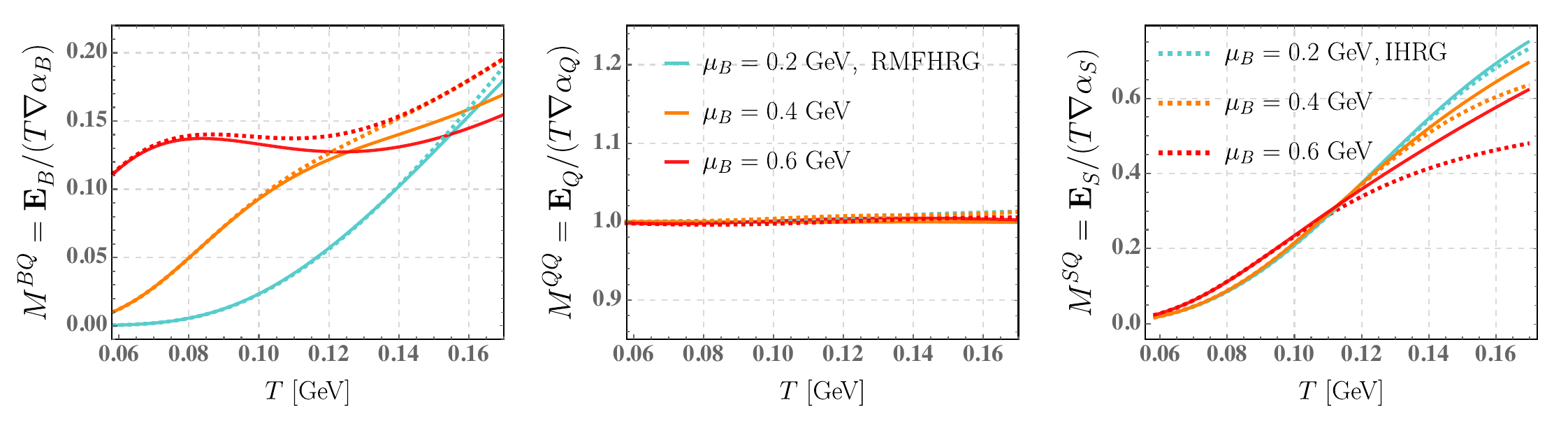}}
\caption{The temperature dependence of conserved charge diffusion thermopower matrix $M^{qQ}$  at $\mu_{B}=$~0.2 GeV (green), 0.4 GeV (orange), 0.6~GeV (red) in the IHRG model (dashed lines) and RMFHRG model (solid lines). The total induced electric field is presented as $\bm{ E}^{}=\bm{ E}^{}_B+\bm{ E}^{}_Q+\bm{E}^{}_S$ with  $\bm{E}_{q}$ being the induced electric field by the gradient in conserved charge chemical potential $\mu_q$.}
\label{fig_M}
\end{figure*}
%-----------------------------------------------------------------------------------
In contrast, the RMF correction significantly enhances  $\sigma^{QS}/T$ and $\sigma^{SS}/T$ at $\mu_{B}=0.6$~GeV. This enhancement is because the predominant carriers for both  $\sigma^{QS}/T$ and $\sigma^{SS}/T$ at $\mu_{B}=0.3$~GeV are still kaons, the RMF interaction between mesons affects the kaon density, while the relaxation time of kaons is influenced by the RMF interaction among various hadron-hadron pairs due to colliding with different hadrons. These effects seem to counteract each other at $\mu_{B}=0.3$~GeV. However, as $\mu_{B}$ rises, the increase in kaon relaxation time caused by the RMF correction significantly overtakes the decrease in kaon density, leading to a substantial elevation in both conductivities at $\mu_{B}=0.6$~GeV. For purely electric conductivity $\sigma^{QQ}/T$, it is nearly unaffected by the RMF correction even at high values of $\mu_B$. 
The reason behind this is that the  RMF correction on the meson and baryon contributions to $\sigma^{QQ}/T$  exactly cancel each other out. 
Overall, the $\mu_{B}$ and $T$ dependence of all conductivities is unaltered by introducing RMF interactions.
%-----------------------------------------------------------------------------------
\begin{figure*}[htpb]
	\centering
	\subfloat{\includegraphics[scale=0.43]{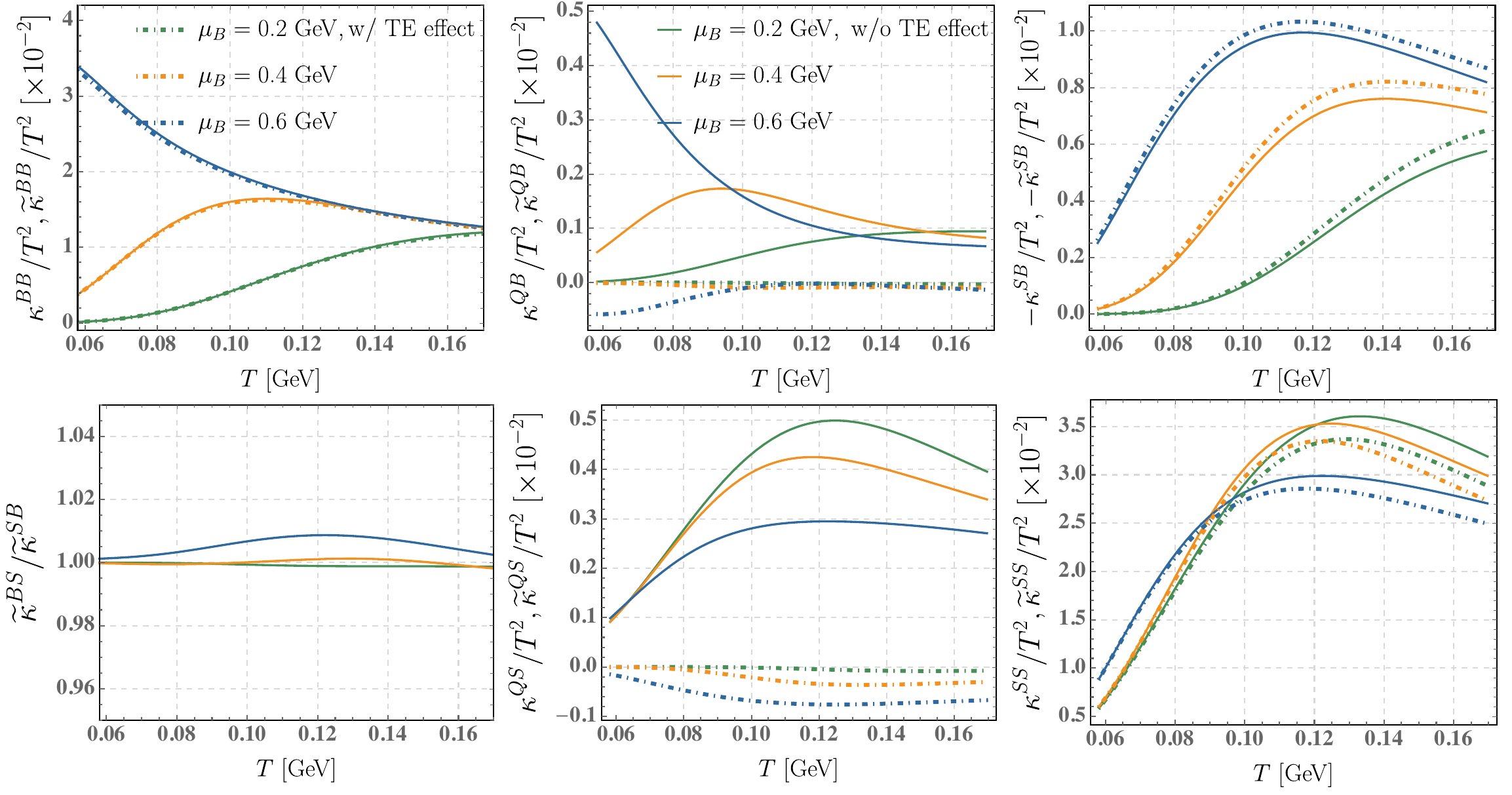}}
	\caption{The comparison between scaled diffusion coefficients matrix $\kappa^{qq''}/T^2$ (solid lines) and scaled thermoelectric modified diffusion coefficient matrix $\widetilde{\kappa}^{qq''}/T^2$ (dot-dashed lines) in the RMFHRG model at $\mu_{B}=$~0.2 GeV (green), 0.4 GeV (orange), 0.6~GeV (blue). The below left panel represents the degree of asymmetry between $\widetilde{\kappa}^{BS}$ and $\widetilde{\kappa}^{SB}$.}
	\label{fig_modify_Kappa}
\end{figure*}

Armed with the knowledge presented above, we can easily comprehend the results of the diffusion coefficient matrix. In Fig.~\ref{fig_Kappa}, we display the $T$ and $\mu_B$ dependence of the complete scaled diffusion coefficient matrix, $\kappa^{qq'}/T^2$, within both the IHRG and RMFHRG models. Akin to the conductivity matrix, the diffusion coefficient matrix elements $\kappa^{qq'}$ and $\kappa^{q'q}$ exhibit symmetry. The integrand of Eq.~(\ref{eq:Kappa}) decomposes the diffusion coefficient matrix element into $\sum_a[q_aq'_a+\widetilde{\epsilon}_a^2\frac{n_qn_q'}{\omega^2}+q_a'\widetilde{\epsilon}_a\frac{n_q}{\omega}+q_a\widetilde{\epsilon}_a\frac{n_q'}{\omega}]\bar{f}_a^0(1\pm \bar{f}_a^0)$.
Within the considered $T$ and $\mu_{B}$ region,  the dominant values of $\widetilde{\epsilon}_a$ in Eq.~(\ref{eq:Kappa}) are lower than $\omega/n_q$.  
Consequently, the qualitative behaviors of the scaled diffusion coefficient matrix are similar to the corresponding scaled conductivity matrix. From Fig.~\ref{fig_Kappa}, we see that the off-diagonal elements can reach a magnitude comparable to the diagonal terms.  Our results within the IHRG model align quantitatively and qualitatively with the results obtained by A.~Das {\it et al}.~\cite{Das:2021bkz} (for a more detailed analysis, see the Appendix).
We note that the values of $\kappa^{BB}/T^2$ and $\kappa^{QB}/T^2$ for high $T$ is larger at $\mu_{B}=0.4$~GeV  than at $\mu_{B}=0.6$~GeV in the IHRG model.  As the temperature rises sufficiently, $\kappa^{QB}/T^2$ at $\mu_{B}=0.2$~GeV can even surpass that at $\mu_{B}=0.4$~GeV. The decreasing dependence of  $\kappa^{BB}$ and $\kappa^{QB}$ on $\mu_{B}$ in high $T$ regions is consistent with the results in the QGP within dynamical quasi-particle model~\cite{Fotakis:2021diq} and holographic model~\cite{Rougemont:2015ona, Grefa:2022sav}. Strikingly different from the $\sigma^{BB}/T$, the $\kappa^{BB}/T^2$ is nearly  unaffected by the RMF interactions at $\mu_B=0.6~$GeV, as depicted in Fig.~\ref{fig_Kappa}. This phenomenon can be 
 attributed to the near cancellation between the decrease in $\sigma^{BB/T}$ and the increase in the integral of $\sum_a\widetilde{\epsilon}_a^2n_B^2/(\omega^2T^2)\bar{f}_a^0(1\pm \bar{f}^0_a)$ caused by the RMF correction in Eq.~(\ref{eq:Kappa}).
 In Fig.~\ref{fig_Kappa}, it is also evident that the RMF correction can increase $-\kappa^{BS}/T^2$ at $\mu_{B}=0.6$~GeV, which contrasts with the trend of $-\sigma^{BS}/T^2$ at $\mu_{B}=0.6$~GeV, shown in Fig.~\ref{fig_electric_conductivity}. This can be well understood from Eq.~(\ref{eq:Kappa}), in which the integral term related to  $-\sum_aS_a\widetilde{\epsilon}_an_B/(\omega T^2)\bar{f}_a^0(1\pm \bar{f}^0_a)$ is enhanced by the inclusion of RMF correction, and this enhancement can overwhelm the reduction in $-\sigma^{BS}$ caused by the RMF correction, thereby leading to an enhancement in $-\kappa^{BS}/T^2$.  It is worth mentioning that, although in the statement of Refs.~\cite{Fotakis:2019nbq, Greif:2017byw, Fotakis:2021diq}, the matching condition in the local rest frame is imposed during the derivation of diffusion coefficient matrix, the obtained formula bear similar to the $\eta^{qq'}T$ given in Eq.~(\ref{eq:eta}) excluding the quantum statistic effect and repulsive mean-field effect (for a detailed discussion, see the Appendix).

Until now, our analysis confines the scenario where the gradients of conserved charge densities are completely directly converted to diffusion currents. However, as mentioned in the introduction, these gradients can also generate an electric field, subsequently influencing the diffusion currents of conserved charges. To exhibit the response of the electric field to gradients in conserved charge chemical thermal potentials,  we display the variations in the  diffusion thermopower matrix $M^{qQ}$ with respect to $T$ and $\mu_{B}$ within both the IHRG and RMFHRG models. As depicted in Fig.~\ref{fig_M},  both $M^{BQ}$ and $M^{SQ}$ are increasing functions with $T$, whereas the $M^{QQ}$  remains almost unchanged and approaches 1. This behavior of $M^{QQ}$ can well be understood from Eq.~(\ref{eq:MqQ}), wherein $\kappa^{QQ}/T^2$ is nearly equivalent in magnitude  to $\eta^{QQ}/T$. Notably, $M^{BQ}$ exhibits the smallest values, indicating a weak ability of the hadron gas (primarily protons) to convert the gradient of baryon chemical thermal potential into an electric field. With increasing $\mu_{B}$, we observe a visible increase in  $M^{BQ}$, while at high $T$, the values of $M^{BQ}$ for different $\mu_{B}$ tend to converge. The decrease of $M^{SQ}$ with respect to $\mu_{B}$ in high $T$ regime is attributed that the strong dependence of $\kappa^{SQ}/T^2$  on $\mu_B$ dominates over the decreasing trend of $T/\eta^{QQ}$ with respect to $\mu_{B}$. Furthermore, as evident from Fig.~\ref{fig_M}, the RMF correction notably augments  $M^{SQ}$ and diminishes $M^{BQ}$ at high $\mu_{B}$. These responses  primarily originate from $\kappa^{SQ}$ and $\kappa^{BQ}$, respectively.

%-----------------------------------------------------------------------------------
\begin{figure*}[htpb]
%\centering
\subfloat{\includegraphics[scale=0.43]{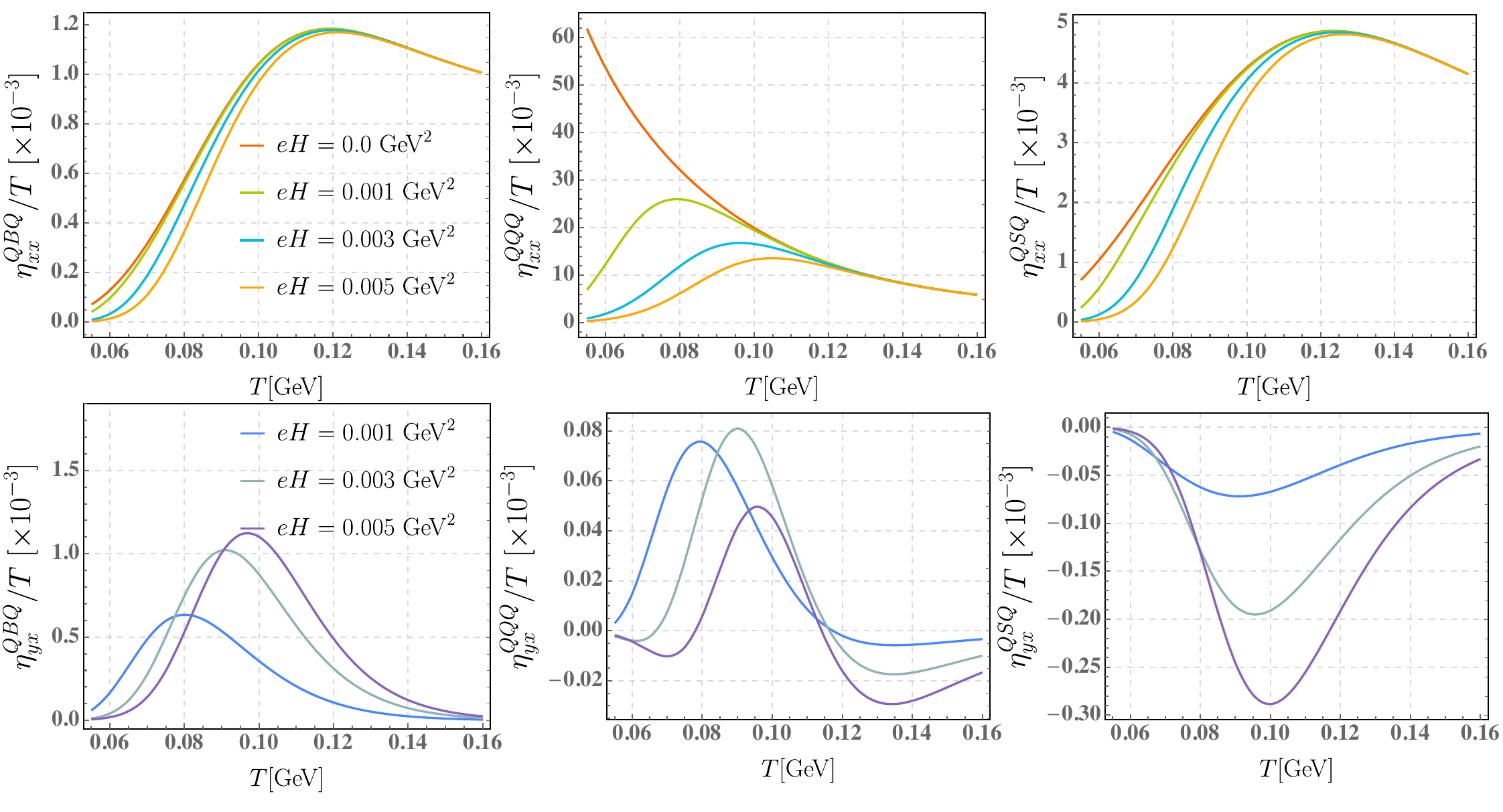}}
\caption{Upper panel: the scaled magnetic field-dependent thermoelectric conductivity matrix $\eta_{xx}^{QqQ}/T$ as a function of temperature in the RMFHRG model for different values of the magnetic field, i.e., $eH=0.0~\mathrm{GeV^2}$ (warm red), $eH=0.001~\mathrm{GeV^2}$ (warm green), $eH=0.003~\mathrm{GeV^2}$ (warm blue), $eH=0.005~\mathrm{GeV^2}$  (warm orange) at $\mu_{B}=0.3$~GeV. Lower panel: the scaled Hall-like thermoelectric conductivity matrix $\eta_{yx}^{QqQ}/T$ as a function of temperature in the RMFHRG model for  $eH=0.001~\mathrm{GeV^2}$ (cool blue),~$eH=0.003~\mathrm{GeV^2}$ (cool green),~$eH=0.005~\mathrm{GeV^2}$ (cool purple) at $\mu_{B}=0.3$~GeV.}\label{fig_eta_B}
\end{figure*}

To intuitively illustrate the impact of the thermoelectric effect on the diffusion coefficient matrix, a comparison between $\kappa^{qq''}/T^2$ and  $\widetilde{\kappa}^{qq''}/T^2$ within the RMFHRG model is given in Fig.~\ref{fig_modify_Kappa}. When the thermoelectric effect is considered, diffusion coefficients in the electric current sector vanish, only those in the baryon and strangeness current sectors exist. Let's first explore the diffusion coefficients in the baryon current sector. From the upper panel of Fig.~\ref{fig_modify_Kappa},  it's evident that $\kappa^{BB}$ is nearly unaffected by the thermoelectric effect. This observation can be well understood from Eq.~(\ref{eq:kappaqq'}), where the product of $M^{BQ}$ and $\eta^{BQ}$ is comparatively small compared to $\kappa^{BB}$.
Different from ${\kappa}^{BB}/T^2$, the inclusion of the thermoelectric effect  obviously diminishes ${\kappa}^{SS}/T^2$.
Additionally, the thermoelectric effect enhances $-\kappa^{SB}/T^2$, which further reduces the net baryon current. Remarkably, the thermoelectric effect  significantly suppresses $\kappa^{QB}/T^2$ ($\kappa^{QS}/T^2$), making $\widetilde{\kappa}^{QB}$ ($\widetilde{\kappa}^{QS}$) much smaller than $\kappa^{BQ}$ ($\kappa^{SQ}$) and even altering its sign. These findings indicate that the thermoelectric effect significantly weakens the correlation between electric current and baryon (strangeness) current, rendering the gradient of electric chemical potential insignificant for baryon (strangeness) diffusion. As a consequence, the baryon diffusion current can be decreased by the thermoelectric effect, resulting in $V_B\approx\widetilde{\kappa}^{BB}\nabla\alpha_B+\widetilde{\kappa}^{BS}\nabla\alpha_S$ if gradients are of comparable magnitudes. As depicted in the lower left panel of Fig.~\ref{fig_modify_Kappa}, the inclusion of thermoelectric effect does not explicitly break the symmetry between $\widetilde{\kappa}^{SB}$ and $\widetilde{\kappa}^{BS}$. 
We remark that the $T$ and $\mu_{B}$ dependence of $\widetilde{\kappa}^{qq''}/T^2$ is still consistent with that of $\kappa^{qq''}/T^2$.   

%-----------------------------------------------------------------------------------
\begin{figure*}[htpb]
\centering
\subfloat{\includegraphics[scale=0.43]{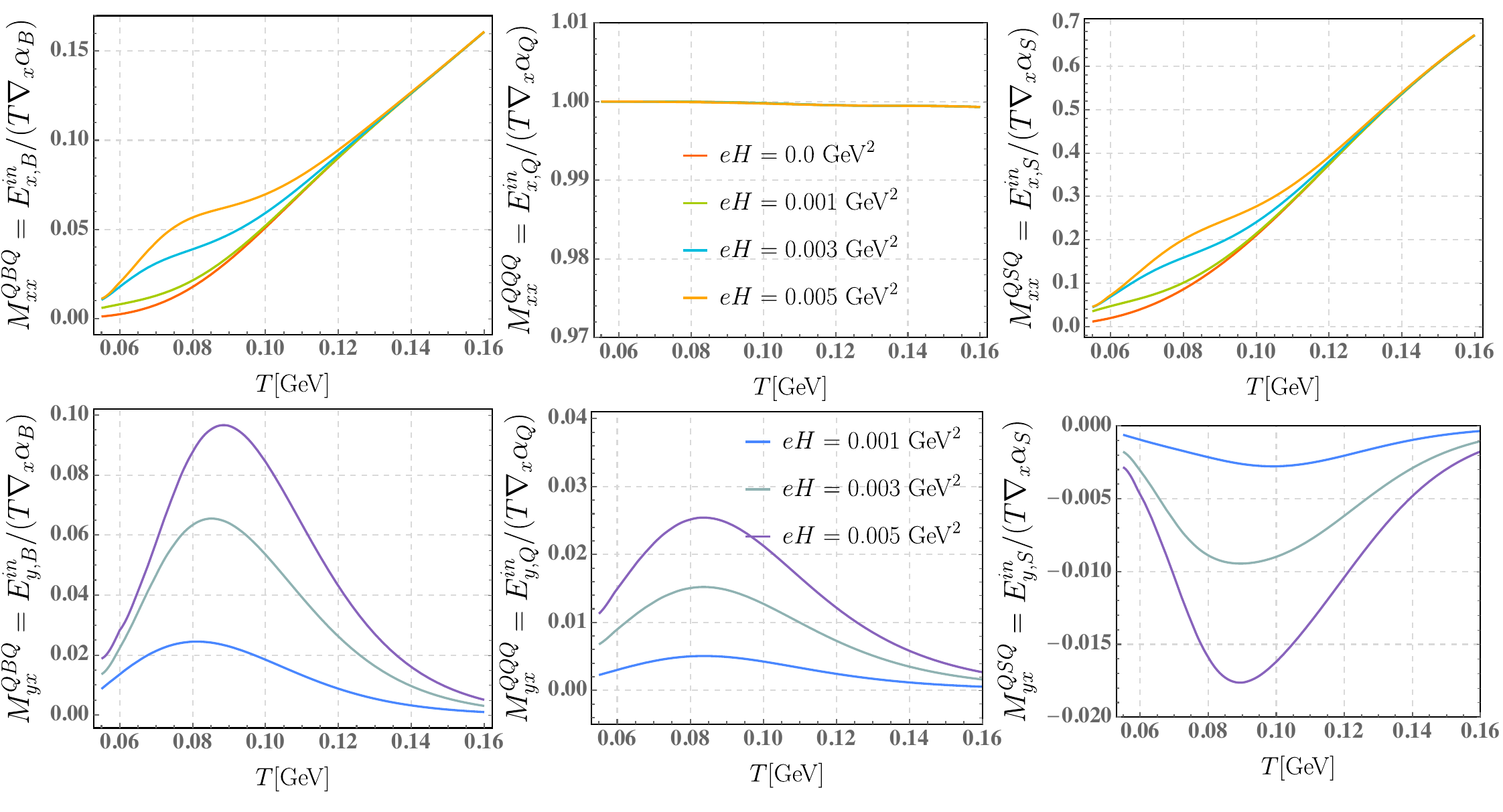}}
\caption{Same as Fig.~\ref{fig_eta_B} but for magnetic field-dependent diffusion thermopower matrix $M_{xx}^{QqQ}$ and Hall-like diffusion thermopower matrix $M_{yx}^{QqQ}$.}
\label{fig_M_B}
\end{figure*} 
%-----------------------------------------------------------------------------------
\subsection{Results for finite magnetic field}
We also study the magneto-thermoelectric effect of hadronic matter  
and examine its impact on both the magnetic field-dependent diffusion coefficient matrix ($\kappa_{xx}^{Qqq'}$) and the Hall-like diffusion coefficient matrix ($\kappa_{yx}^{Qqq'}$).
So far, the realistic time evolution of the initial magnetic field remains unclear. Based on the simple parametrization $eH(\sqrt{s_{NN}})= 0.021\sqrt{s_{NN}}m_{\pi}^2$ ($m_{\pi}$ is the pion mass) for Au+Au collisions with fixed collision parameter $b=10~\mathrm{fm}$~\cite{Deng:2012pc}, and using $eH=eH_0(\tau_0/\tau)^a$ with $a=1$ as well as  $eH\sim 4m_{\pi}^2$ for the thermalization timescale $\tau_0\sim1$~fm, we estimate that for the hadronization time scale $\tau\sim 10$~fm, $eH$ would be approximately  $ 0.008~\mathrm{GeV^2}$. In our study,  we consider a magnetic field region ranging from 0 to  $0.005~\mathrm{GeV^2}$, as done in Ref.~\cite{Das:2020beh}. 

All the estimations at finite magnetic fields are performed using the RMFHRG model, with a fixed baryon chemical potential $\mu_{B}=0.3$~GeV. As illustrated in the upper panel of Fig.~\ref{fig_eta_B}, the temperature dependence of the scaled magnetic field-dependent thermoelectric conductivity, specifically  $\eta^{Qq''Q}_{xx}/T$ where $q''\in \{B,S\}$, remains unaltered at finite magnetic fields when compared to zero magnetic field. It is observed that $\eta_{xx}^{QQQ}/T$ at finite magnetic fields first increases with temperature and subsequently decreases. This non-monotonic behavior of $\eta_{xx}^{QQQ}/T$ with respect to $T$ is primarily attributed to the combined effect of the magnetic field and relaxation time in the integrand of Eq.~(\ref{eq:eta-xx-xy}), which can be well understood in analogy with the discussion on magnetic field-dependent electric conductivity reported in  previous studies~\cite{Das:2019wjg, Zhang:2020efz,Das:2019ppb}. The dominant charge carriers for $\eta_{xx}^{QQQ}/T$ are charged pions. At low $T$, the scattering rate of pions is smaller than cyclotron frequency ($\omega_{c}$), resulting in  $\frac{\tau_a}{(\omega_{c,a}\tau_a)^2+1}\sim 1/(\omega_{c,a}^2\tau_{a})$. While, at high $T$, the pion scattering rate exceeds $\omega_{c}$, leading to  $\frac{\tau_a}{(\omega_{c,a}\tau_{a})^2+1}\sim \tau_{a}$.
Furthermore, all the scaled magnetic field-dependent thermoelectric conductivities initially decrease with increasing magnetic field before converging at high $T$.
From the lower panel of Fig.~\ref{fig_eta_B},  we also note that all scaled Hall-like thermoelectric conductivities, denoted as $\eta_{yx}^{QqQ}/T$, exhibit a non-monotonic temperature dependence. This behavior is attributed to the complex interplay of various factors, including relaxation time, cyclotron frequency, and the factor $n_q/\omega$.
In the hadron gas, the predominant contributions for $\eta_{yx}^{QqQ}/T$ stem from protons (p) and Sigma baryons ($\Sigma^{+}$). 
We note that $\eta_{yx}^{QSQ}/T$ takes on negative values. This can be understood through Eq.~(\ref{eq:eta-xx-xy}), where the dominant  term  in the integrand, $\sim \sum_a S_aQ_a\widetilde{\epsilon}_an_Q/\omega\bar{f}^0_a(1+\bar{f}^0_a)$, determine the sign of $\eta_{yx}^{QSQ}/T$. This sign is predominantly influenced by Sigma baryons. 
The dependence of $\eta^{QBQ}_{yx}/T$ on the magnetic field is non-monotonic. This is because, the predominant proton scattering rate in low $T$ region is much smaller than the corresponding $\omega_{c,a}$,  resulting in $\frac{\omega_{c,a}\tau_a^2}{(\omega_{c,a}^2\tau_{a})^2+1}\sim 1/ \omega_{c,a} $. While in high $T$ regions, the proton scattering rate plays a dominant role with $\omega_{c,a}\tau_{a}\ll 1$, leading to $\frac{\omega_{c,a}\tau_a^2}{(\omega_{c,a}\tau_{a})^2+1}\sim \omega_{c,a}$. In contrast, the dependence of $\eta^{QSQ}_{yx}/T$  on the magnetic field appears almost monotonic since the scattering rate of predominant Sigma baryons is always larger than $\omega_{c,a}$ in the entire  $T$ region considered.   Additionally, we note the magnetic field dependence of  $\eta^{QQQ}_{yx}/T$ is akin to that of $\eta^{QBQ}_{yx}/T$, but its sign differs at low and high $T$. This variance is attributed to the shifting predominant contribution from protons at lower $T$ and  Sigma baryons at higher $T$.
\begin{figure*}[htpb]
	%\centering
	\subfloat{\includegraphics[scale=0.45]{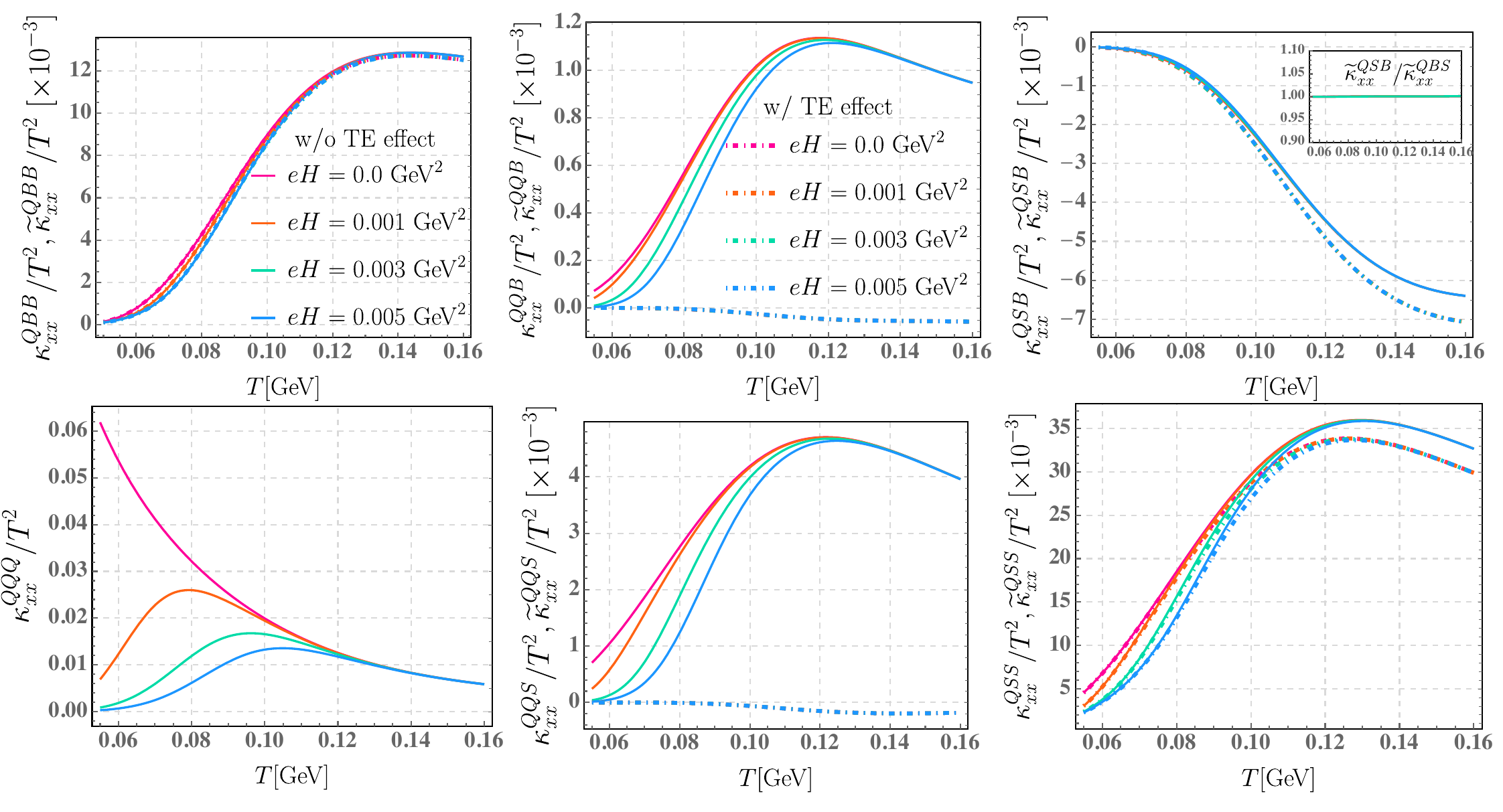}}
	\caption{Both complete scaled magnetic field-dependent diffusion coefficient matrix $\kappa_{xx}^{Qqq'}/T^2$(solid lines) and  complete scaled magneto-thermoelectric modified diffusion coefficient matrix $\widetilde{\kappa}_{xx}^{Qqq''}/T^2$ (dot-dashed lines)  for $eH=0.0~\mathrm{GeV^2}$ (magenta), $0.001~\mathrm{GeV^2}$ (red),~$0.003~\mathrm{GeV^2}$ (green),~$0.005~\mathrm{GeV^2}$ (blue) at $\mu_{B}=0.3$~GeV in the RMFHRG model. The inset in the upper right panel quantifies the degree of asymmetry between $\widetilde{\kappa}_{xx}^{QBS}$ and $\widetilde{\kappa}_{xx}^{QSB}$.}
	\label{fig_X_Kappa}
\end{figure*}
%----------------------------------------------------------------------------------- 
\begin{figure*}[htpb]
\centering
\subfloat{\includegraphics[scale=0.45]{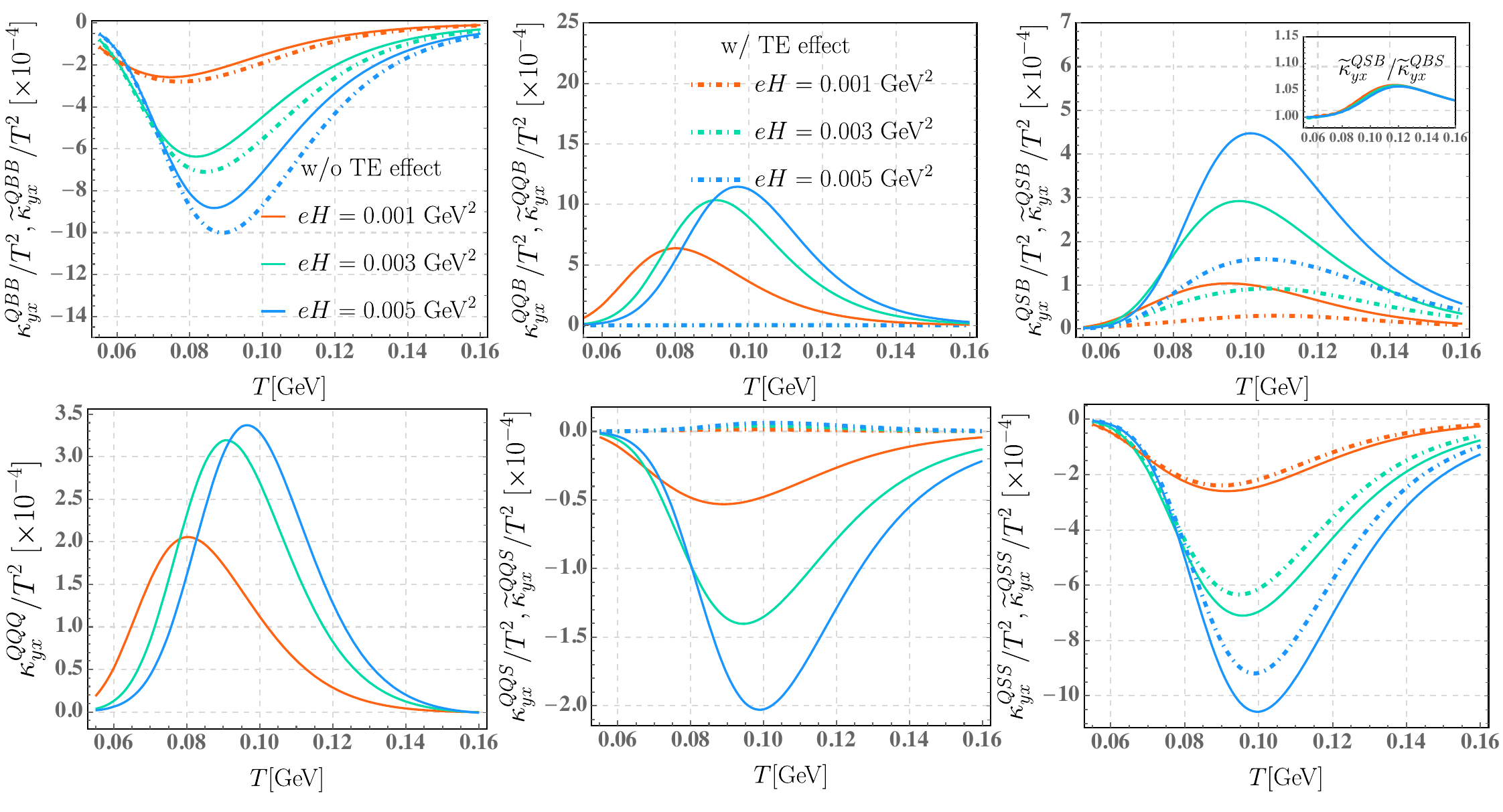}}
\caption{Same as Fig.~\ref{fig_X_Kappa} for both complete scaled Hall-like diffusion coefficient matrix $\kappa_{yx}^{Qqq'}/T^2$ and complete scaled magneto-thermoelectric modified Hall-like  diffusion coefficient matrix $\widetilde{\kappa}_{yx}^{Qqq''}/T^2$.
}
\label{fig_Y_Kappa}
\end{figure*} 
% \FloatBarrier
%-----------------------------------------------------------------------------------
\begin{figure*}[htpb]
\centering
\subfloat{\includegraphics[scale=0.45]%{Fig_X_Y_M_QqQ_iso_adia_U3_1110.pdf}}
{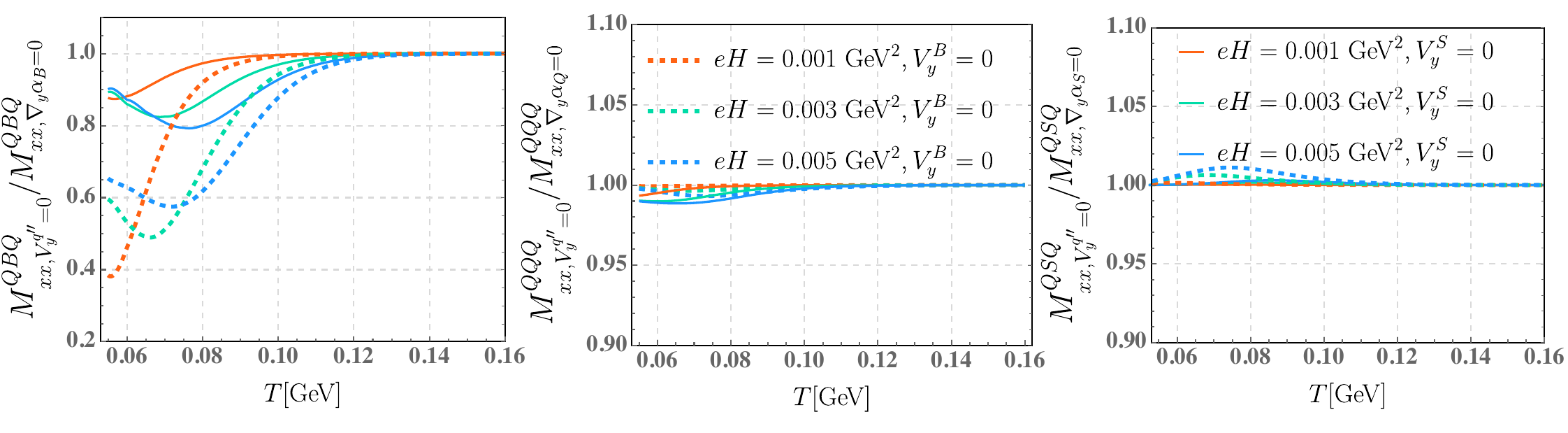}}
\caption{ %(Upper panel)
The ratio of magnetic field-dependent diffusion thermopower
under the condition of $V_y^{q''}=0$  to that 
under the condition of $\nabla_y\alpha_q=0$  as a function of temperature at $eH=0.001~\mathrm{GeV^2}$ (red),~$0.003~\mathrm{GeV^2}$ (green),~$0.005~\mathrm{GeV^2}$ (blue). The dashed lines and solid lines correspond to the results under the condition of $V_y^B=0$ and the condition of $V_y^S=0$, respectively. All the numerical calculations are performed at $\mu_{B}=0.3$~GeV in the RMFHRG model. 
}
\label{fig_X_Y_Ratio_M}
\end{figure*}

Next, we discuss the qualitative behaviors of both magnetic field-dependent diffusion thermopower matrix, $M_{xx}^{QqQ}$, and Hall-like diffusion thermopower matrix, $M_{yx}^{QqQ}$, with respect to temperature and magnetic field, respectively. As shown in the upper panel of Fig.~\ref{fig_M_B}, the application of a magnetic field enhances both $M^{QBQ}_{xx}$ and $M^{QSQ}_{xx}$ in low $T$ region, which means the ability of the hadron gas to convert baryon and strangeness chemical potential gradients into an electric field is strengthened by adding a magnetic field. Whereas the $M^{QQQ}_{xx}$ appears to be independent of the magnetic field. All components of  $M_{yx}^{QqQ}$  exhibit a significant dependence on the magnetic field in the studied temperature region. As seen in the lower panel of Fig.~\ref{fig_M_B}, all $M_{yx}^{QqQ}$ components in magnitude display a similar peak structure throughout the entire $T$ region, and these magnitudes increase as $eH$ increases. It's worth noting that the magnitude of $M_{yx}^{QBQ}$ is comparable to that of  $M_{xx}^{QBQ}$ at low $T$. 

As in the case without a magnetic field, we also compare the magnetic field-dependent diffusion coefficient matrix before and after accounting for the magneto-thermoelectric effect. As illustrated in Fig.~\ref{fig_X_Kappa},
we note that all components of  $\kappa^{Qqq'}_{xx}/T^2$, except for $\kappa^{QQQ}_{xx}/T^2$, remain qualitatively unchanged in the presence of a magnetic field. The associated explanation of $\kappa^{QQQ}_{xx}/T^2$ mirrors that of $\eta^{QQQ}_{xx}/T$. Within the baryon current sector, $\kappa^{QBB}_{xx}$ and  $\kappa^{QSB}_{xx}$  appear insensitive to the magnetic field, while the remaining terms exhibit significant reductions in the low $T$ region when the magnetic field is introduced. Thus, the presence of a magnetic field can significantly hinder the strangeness and electric diffusion, especially when gradients of similar magnitudes are involved.
The qualitative behavior of the magneto-thermoelectric modified diffusion coefficient matrix, $\widetilde{\kappa}^{Qqq''}_{xx}$, aligns with that of the non-modified matrix, ${\kappa}^{Qqq''}_{xx}$. Furthermore, as depicted in the inset of the upper right panel of Fig.~\ref{fig_X_Kappa},  the symmetry between $\widetilde{\kappa}^{QSB}_{xx}$ and $\widetilde{\kappa}^{QBS}_{xx}$  is almost maintained.

In comparison to $\kappa^{Qqq'}_{xx}/T^2$, all Hall-like diffusion coefficient matrix elements, $\kappa^{Qqq'}_{yx}/T^2$,  exhibit a strong dependence on the magnetic field, as shown in Fig.~\ref{fig_Y_Kappa},  and display a peak structure in magnitude. The explanation for the behavior of  $\kappa^{QqQ}_{yx}/T^2$ with respect to  $T$ and $eH$ is akin to that of $\eta^{QqQ}_{yx}/T^2$. 
It's worth noting that $\kappa^{Qqq''}_{yx}/T^2$ show obvious responses to the magneto-thermoelectric effect.
The magneto-thermoelectric effect can considerably diminish the magnitude of $\widetilde{\kappa}_{yx}^{QQB}$ ($\widetilde{\kappa}_{yx}^{QQS}$), thereby rendering the transverse coupling between electric charge and baryon (strangeness) charge insignificant in Hall-like baryon (strangeness) current. 
Strikingly different from $\kappa_{xx}^{QBB}$, we observe that the magneto-thermoelectric effect gives an obvious enhancement in the magnitude of $\kappa_{yx}^{QBB}$. The $eH$ and $T$ dependence of $\widetilde{\kappa}_{yx}^{Qqq''}$ remains similar to that of $\kappa_{yx}^{Qqq''}$. Notably, the introduction of the magneto-thermoelectric effect causes a significant asymmetry between $\widetilde{\kappa}_{yx}^{QSB}$ and  $\widetilde{\kappa}_{yx}^{QBS}$, as depicted in the inset of the upper right panel of Fig.~\ref{fig_Y_Kappa}. This asymmetry intensifies further with the increase in $\mu_{B}$.

The aforementioned calculations regarding magnetic field-dependent diffusion thermopower matrix and magneto-thermoelectric modified diffusion coefficient matrix in the presence of a magnetic field were performed under the assumption that the conserved charge chemical thermal potential gradients are solely along the longitudinal direction, specifically with, $\nabla_x\alpha_q\neq0,~\nabla_y\alpha_q=0$. As stated in Sec.~\ref{sec:coefficient}, under the condition of zero transverse diffusion current of conserved charge, (i.e., $V_y^{q''}=0$ with $q''\in\{B,S\}$),  a transverse gradient $\nabla_y\alpha_q$ can arise from $\nabla_x\alpha_q$, subsequently inducing a transverse electric field. Under these conditions, the magnitude of the magneto-thermoelectric modified diffusion coefficient matrix might differ from that under the condition of $\nabla_y\alpha_q=0$. To intuitively quantify the impact of varying transverse conditions,  Fig.~\ref{fig_X_Y_Ratio_M}  illustrate the ratio of magnetic field-dependent diffusion thermopower computed under the conditions of $V_y^{B}=0$ and $V_y^{S}=0$ to that determined under the condition of $\nabla_y \alpha_q=0$. 
We find that imposing the condition of $V_y^{q''}=0$ results in an obvious reduction in $M^{QBQ}_{xx}$ within low $T$ region, whereas both $M_{xx}^{QQQ}$ and $M_{xx}^{QSQ}$ exhibit almost insensitive to alterations in transverse conditions.
This result is unsurprising since, under the condition of $\nabla_y\alpha_{q}=0$, the values of $M_{yx}^{QQQ}$ ($M_{yx}^{QSQ}$) is much smaller than the corresponding $M_{xx}^{QQQ}$ ($M_{xx}^{QSQ}$), as shown in Fig.~\ref{fig_M_B}. Additionaly, the $\mathcal{L}^{q}_{V_{y}^{q''}=0}$ always remains less than $1$ due to $|\widetilde{\kappa}_{yx}^{Qqq''}|<|\widetilde{\kappa}_{xx}^{Qqq''}|$, consequently, the product of $M_{yx}^{QQQ}$ ($M_{yx}^{QSQ}$) and $\mathcal{L}^Q_{V_y^{q''}=0}$ in Eq.~(\ref{eq:Adia_Mxx}) is negligible in  comparison to  $M_{xx}^{QQQ}$ ($M_{xx}^{QSQ}$).

%-----------------------------------------------------------------------------------
\begin{figure}[htpb]
	\subfloat{\includegraphics[scale=0.34]{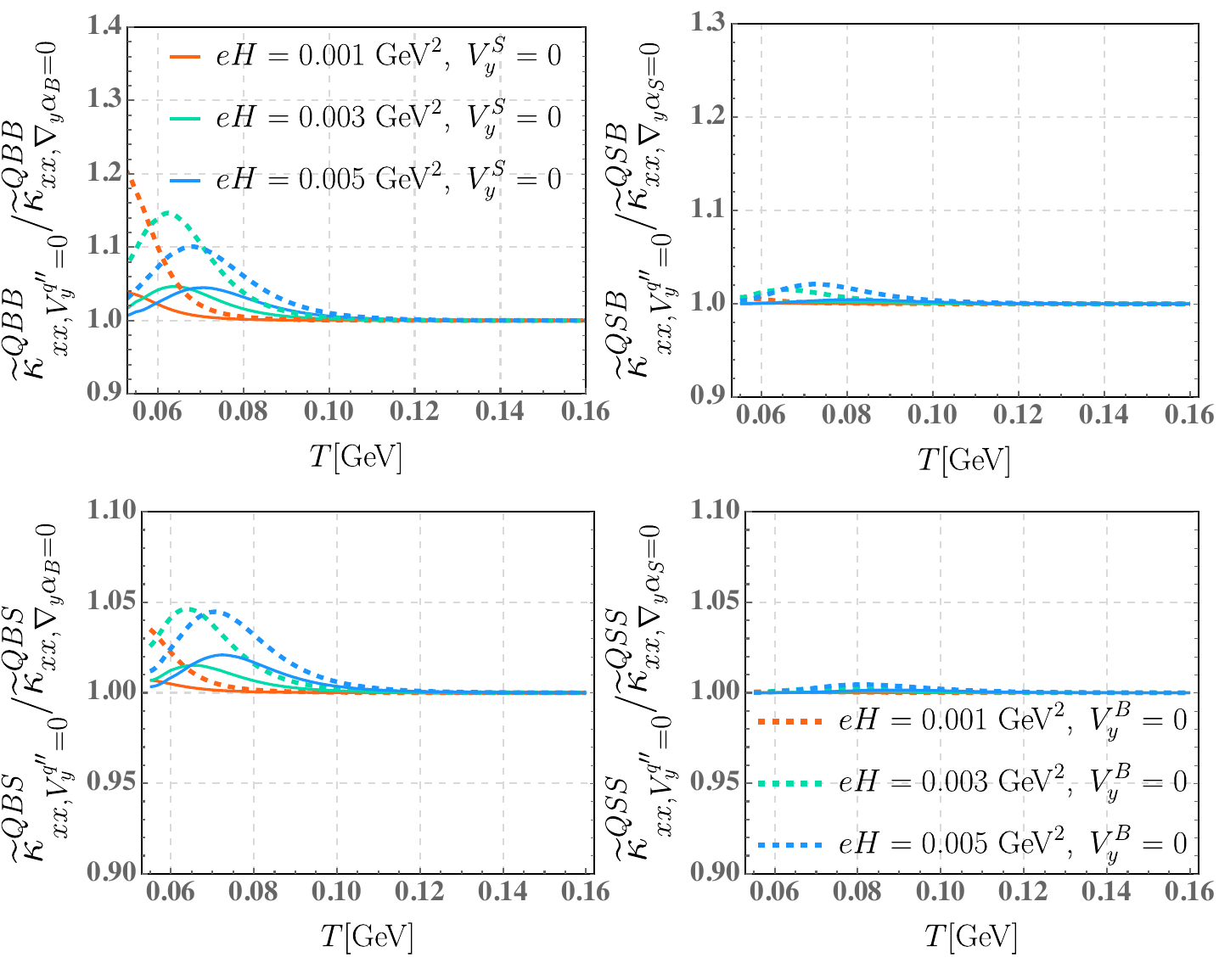}}
	\caption{Same as Fig.~\ref{fig_X_Y_Ratio_M} but for the ratio of magneto-thermoelectric modified diffusion coefficient  under the condition of $V_y^{q''}=0$ to that under the condition of $\nabla_y\alpha_q=0$. }
	\label{fig_Ratio}
\end{figure} 
Finally, we show the sensitivity of magneto-thermoelectric modified diffusion coefficient matrix, $\widetilde{\kappa}_{xx}^{Qqq''}$, to various choices of transverse conditions in Fig.~\ref{fig_Ratio}.  The results for $\widetilde{\kappa}_{xx}^{QQB}$ and $\widetilde{\kappa}_{xx}^{QQS}$ are not presented here as they are much smaller than the other terms. We observe that taking the condition of $\nabla_{y}\alpha_q=0$ as a baseline, the variation of $\widetilde{\kappa}_{xx}^{Qqq''}$ due to changes in transverse conditions is generally small, except that $\widetilde{\kappa}_{xx}^{QBB}$  has a slight reduction in the condition of $V_{y}^B=0$ at low $T$.  The qualitative features of $\widetilde{\kappa}_{xx}^{Qqq''}$  almost remain unchanged  under various transverse 
conditions.

%----------------------------------------------------------------------------------- 
\section{summary}\label{sec:summary}
We investigated the thermoelectric effect and diffusion process involving multiple conserved charges in hot and dense hadronic matter. Their corresponding diffusion thermopower matrix $M^{qQ}$ and diffusion coefficient matrix $\kappa^{qq'}$ with $q,q'\in \{B,Q,S\}$ were evaluated in both the IHRG and RMFHRG models by solving the relativistic Boltzmann equation under relaxation time approximation, where the Landau-Lifshitz energy frame was adopted. In the RMFHRG model, the repulsive interaction between hadrons is treated as a density-dependent mean-field potential, leading to a shift in the single-particle energy. 
In the presence of a magnetic field,  additional Hall-like diffusion thermopower matrix $M_{yx}^{QqQ}$ and Hall-like diffusion coefficient matrix $\kappa_{yx}^{Qqq'}$ emerge.
We further explored the impact of the magneto-thermoelectric effect on both ${\kappa}_{xx}^{Qqq'}$ and ${\kappa}_{yx}^{Qqq'}$. Additionally, we studied the sensitivities 
of magnetic field-dependent diffusion thermopower matrix $M_{xx}^{QqQ}$ and magneto-thermoelectric modified diffusion coefficient matrix $\widetilde{\kappa}_{xx}^{Qqq''}$ (where $q''\in \{B,S\}$) to various transverse restrictions.
Below, we outline the primary findings emerging from our research.
\begin{itemize}
\item 
 All the scaled diffusion coefficients, except for  $\kappa^{QQ}/T^2$ and $\kappa^{BB}/T^2$ are sensitive to the RMF interactions in the baryon-rich region, indicating that the repulsive interactions between hadrons are crucial for understanding the diffusion properties of QCD matter created at the lower collision energies. 
 
\item
Both $M^{BQ}$ and $M^{SQ}$ exhibit a strong dependence on $T$ and $\mu_{B}$. In contrast, the $M^{QQ}$  remains almost unaffected by varying $T$ and $\mu_{B}$, maintaining a value close to 1. %The RMF correction significantly enhances $M^{SQ}$ and reduces $M^{BQ}$ at large $\mu_B$. 
The introduction of RMF corrections leads to a substantial increase in $M^{SQ}$ and a decrease in $M^{BQ}$ at large $\mu_B$. 
 
\item
%The thermoelectric effect generally prevents the baryon (strangeness) diffusion, and in particular, significantly decouples the correlation between electric charge and baryon number (strangeness). 
The thermoelectric effect generally hinders baryon (strangeness) diffusion and significantly weakens the correlation between electric charge and baryon number (strangeness).

\item 
In the magnetic field, both $M_{xx}^{QBQ}$ and $M_{xx}^{QSQ}$ increase with the magnetic field at low $T$, whereas $M_{xx}^{QQQ}$ is almost magnetic field independent. The magnitude of the Hall-like diffusion thermopower matrix is considerably influenced by $eH$ and exhibits a distinct peak structure in the considered $T$ region. Compared to $M_{xx}^{QQQ}$ and $M_{xx}^{QSQ}$, the magnitude of $M_{xx}^{QBQ}$ at low $T$ is more sensitive to variations in transverse restriction conditions.

\item
Apart from $\kappa_{xx}^{QBB}$ and $\kappa_{xx}^{QSB}$, the other diffusion coefficients are sensitive to $eH$ and decrease with $eH$  at low $T$, indicating that the magnetic field can impede the electric charge and strangeness diffusion. Furthermore,
the quantitative and qualitative characteristics of $\widetilde{\kappa}^{Qqq''}_{xx}$ remain relatively stable under varying transverse restriction conditions.

% \hspace{0.3cm}
 
\item 
The full Hall-like diffusion coefficients in magnitude reveal a similar peak structure in the considered $T$ region and exhibit a strong dependence on the magnetic field. Notably, the inclusion of the magneto-thermoelectric effect can result in asymmetry between  $\widetilde{\kappa}_{yx}^{QSB}$ and $\widetilde{\kappa}_{yx}^{QBS}$.  
\end{itemize}
 These findings could offer valuable insights into the dynamics of various conserved charges and contribute to the development of dissipative (magneto-)hydrodynamics frameworks that explicitly incorporate multiple conserved charges.

\acknowledgments
This work was supported by Guangdong Major Project of Basic and Applied Basic Research No. 2020B0301030008 and No. 2022A1515010683, the Natural Science Foundation of China No.12247132, and the China Postdoctoral Science Foundation No.2023M731159.

\section*{appendix}
\begin{figure*}[htpb]
\centering
\subfloat{\includegraphics[scale=0.4]{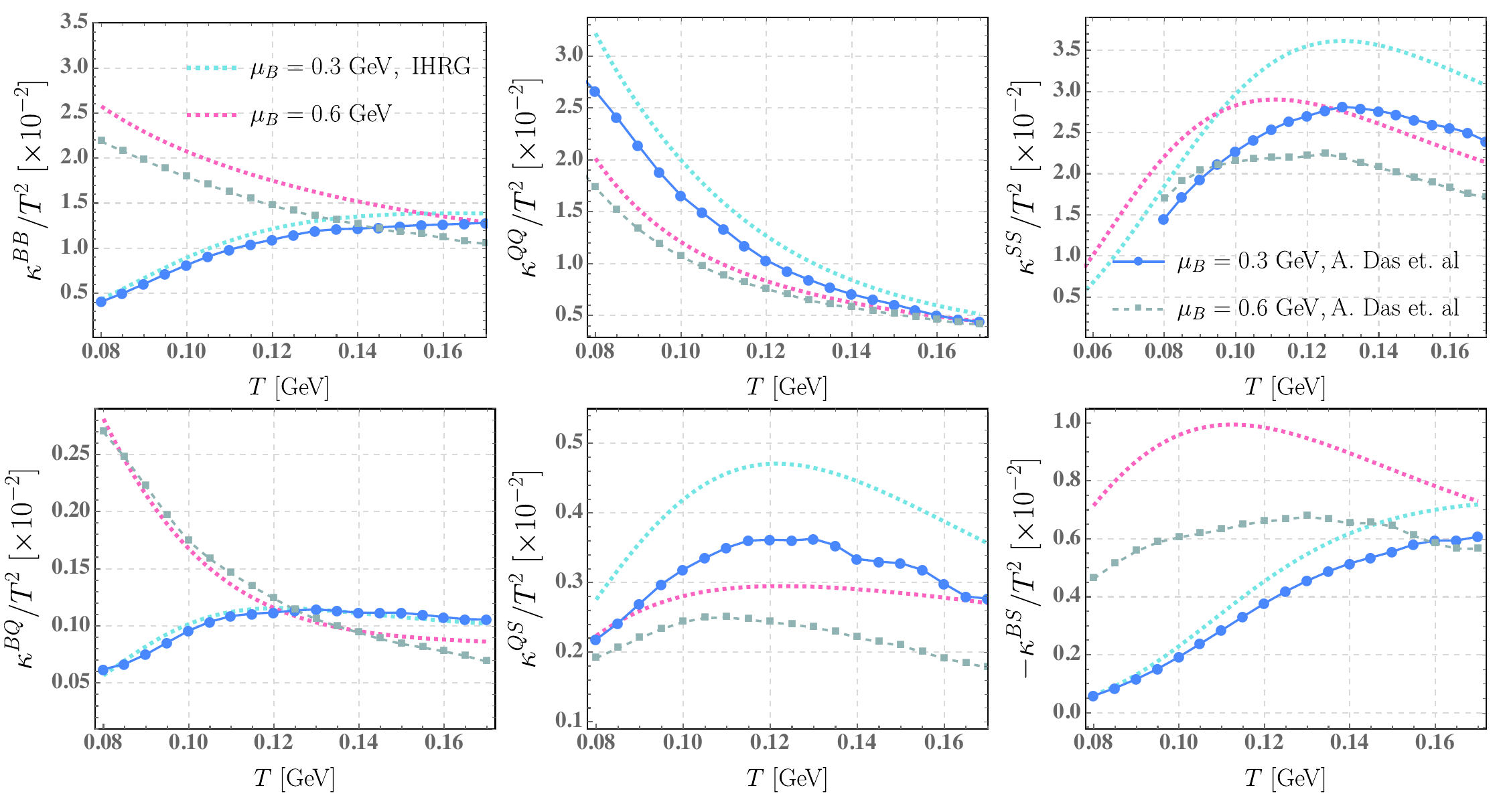}}
\caption{The diffusion coefficient matrix results obtained by A.~Das {\it et al}. (symbol lines)~\cite{Das:2021bkz} is compared to our results (dashed lines) in the IHRG model with the same settings ($n_S=0,~\mu_Q=0$) at $\mu_{B}=0.3$~GeV and  $\mu_{B}=0.6$~GeV. 
}
\label{fig_Kappa_comparison}
\end{figure*}

In Fig.~\ref{fig_Kappa_comparison}, we present a comparison between the diffusion coefficient matrix result obtained in this study and the one reported by A.~Das {\it et al}. in Ref.~\cite{Das:2021bkz} within the framework of the IHRG model. 
The diffusion coefficient matrix $\kappa^{qq'}$ derived in Ref.~\cite{Das:2021bkz} takes the following form: \begin{align}
\kappa^{qq'}=&\sum_a\frac{d_a}{3}\int \frac{d^3p_a}{(2\pi)^3}\tau_a\frac{\bm{p}_a^2}{(\epsilon^0_a)^2}\left[q'_a-\epsilon_a^0\frac{n'_q}{\omega}\right]\nonumber\\
&\times\left[q_a-\epsilon_a^0\frac{n_q}{\omega}\right]f^{(0)}_a,
\end{align}
where $f^{(0)}_a$ represents the equilibrium distribution function in the classical limit, i.e., the Boltzmann distribution function.
As illustrated in Fig.~\ref{fig_Kappa_comparison},  our results for the full diffusion coefficient align closely
with those reported by A.~Das {\it et al}. The numerical discrepancies primarily arise from the choices in degrees of freedom and the absence of quantum statistic effect in~\cite{Das:2021bkz}.

\begin{figure*}
\centering
\subfloat{\includegraphics[scale=0.4]{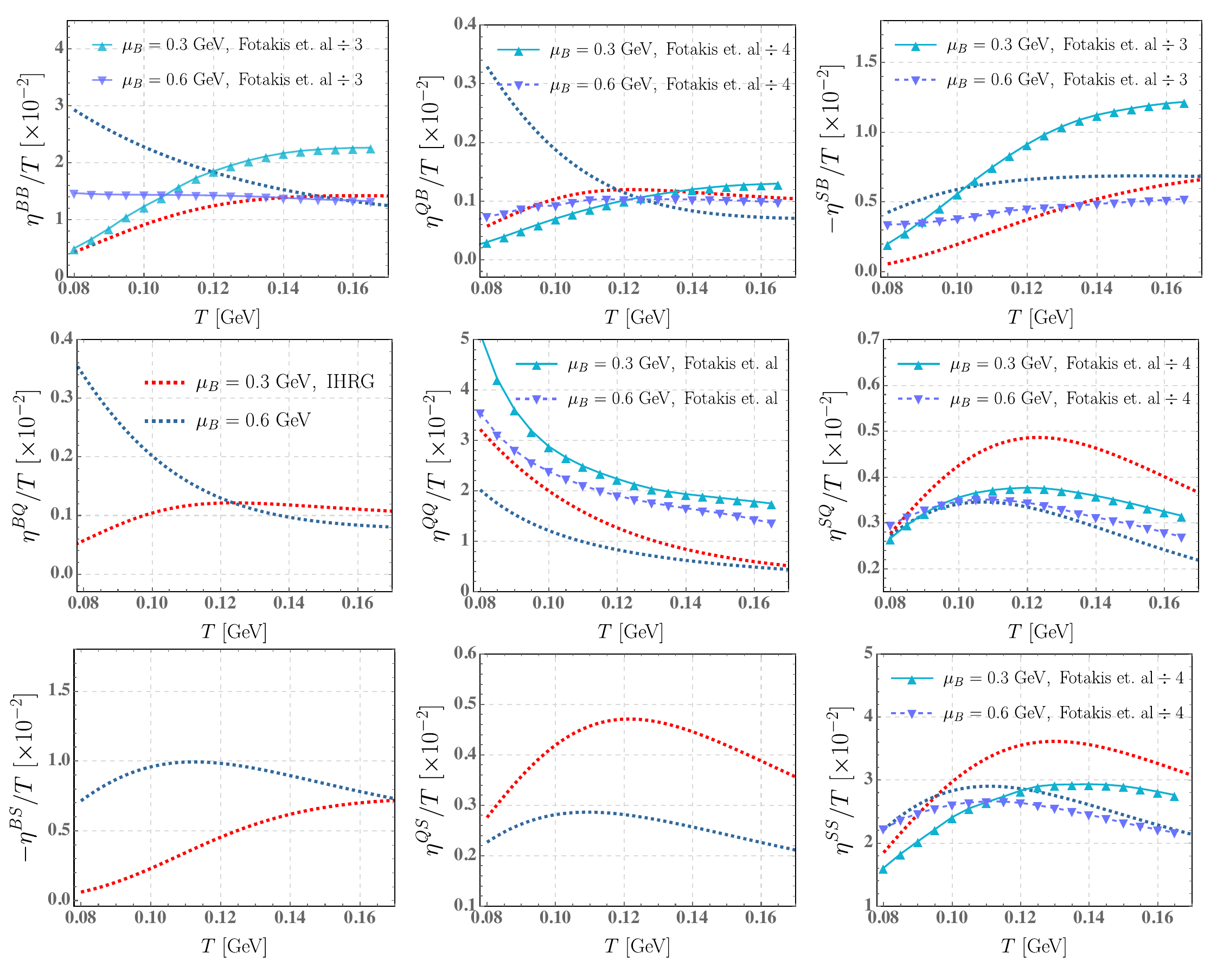}}
\caption{The temperature dependence of the scaled thermoelectric transport coefficient matrix
$\eta^{qq'}/T$ in baryon (top row), electric (middle row), strangeness diffusion current sectors (bottom row) at $\mu_{B}=$~0.3 GeV (red), 0.6~GeV (blue) using the IHRG model (dashed lines). These results are compared with the diffusion coefficient matrix result obtained by Fotakis {\it et al}. at $\mu_B=0.3$~GeV (upper triangles) and 0.6~GeV (lower triangles) with the same settings ($n_S=0,~\mu_Q=0$)~\cite{Fotakis:2019nbq}.
}
\label{fig_Eta}
\end{figure*}

 Alternatively,  we note that the expression of $\kappa^{qq'}$ presented in Refs~\cite{Fotakis:2019nbq, Greif:2017byw, Fotakis:2021diq}  differs from ours. In those references, $\kappa^{qq'}$ is expressed as 
\begin{align}
\kappa^{qq'}=&\frac{\tau}{3}\sum_a\frac{d^3p_a}{(2\pi)^3}\frac{\bm{p_a}^2}{(\epsilon^0_a)^2}(q_a'-\frac{\epsilon^0_an_q'}{\omega})q_af^{(0)}_a.
%\\=&\frac{\tau}{3}\sum_a\frac{d^3p_a}{(2\pi)^3}\frac{\bm{p_a}^2}{(\epsilon^0_a)^2}q_a'q_af^{(0)}_a-\tau\frac{Tn_{q'}n_q}{\omega}.
\end{align}
In Ref.~\cite{Fotakis:2019nbq}, a constant relaxation time $\tau$ for all the hadron species is employed. Comparing with our thermoelectric transport coefficient $\eta^{qq'}$ from Eq.~(\ref{eq:eta}), we find that $\eta^{qq'}/T$  bears formal resemblance to $\kappa^{qq'}/T^2$ derived in Refs.~\cite{Fotakis:2019nbq, Greif:2017byw, Fotakis:2021diq}, discounting the quantum statistic effect and repulsive mean-field effect. As shown in Fig.~\ref{fig_Eta}, our results are smaller than those reported in Ref.~\cite{Fotakis:2019nbq}, and the qualitative behaviors of $\eta^{qq'}/T^2$ in the baryon diffusion current are also slightly different from the results in Ref.~\cite{Fotakis:2019nbq}.
It is clear that $\eta^{BS}/T$ in Fig.~\ref{fig_Eta} is not equivalent to $\eta^{SB}/T$ at $\mu_{B}=0.6$~GeV. This indicates that the symmetry of the off-diagonal diffusion coefficients ($\kappa^{qq'}=\kappa^{q'q}$) reported in Refs.~\cite{Fotakis:2019nbq, Greif:2017byw} may not hold true when the quantum statistic effects are considered. By comparing Fig.~\ref{fig_Kappa} and Fig.~\ref{fig_Eta}, we observe that the quantitative and qualitative differences between $\kappa^{qq'}/T^2$ and $\eta^{qq'}/T$ are negligible. Therefore, the numerical discrepancy between our diffusion coefficient matrix and that of Ref.~\cite{Fotakis:2019nbq} can primarily be attributed to differences in the degrees of freedom and relaxation times. 

%-----------------------------------------------------------------------------------
% \FloatBarrier

%---------------------------
%----------------------------------------------------------------------------------- 
% \FloatBarrier

\end{document}